\documentclass[a4paper,fleqn,usenatbib,useAMS]{mnras}

\usepackage{graphicx}	
\usepackage{amsmath}	
\usepackage{amssymb}	
\usepackage{multicol}        
\usepackage{bm}		
\usepackage{pdflscape}	
\usepackage{journals}
\usepackage{epstopdf}

\newcommand{\Msun}{M\ensuremath{_\odot}\,}

\def\HI{H~{\sc i}\, }
\def\HII{H~{\sc ii}\, }

\def\kms{$\textrm{km~s$^{-1}$}$}
\def\nb{\textsc{nbursts}}

\usepackage[T1]{fontenc}
\usepackage{ae,aecompl}

\usepackage{newtxtext,newtxmath}

\title[A Malin\ 1 ``cousin'' with counter-rotation]{A Malin~1 ``cousin'' with counter-rotation: internal dynamics and stellar content of the giant low surface brightness galaxy UGC\ 1922}
\author[A. Saburova et al.]{
Anna S. Saburova,$^{1}$\thanks{E-mail:saburovaann@gmail.com}
Igor V. Chilingarian,$^{2,1}$
Ivan Yu. Katkov,$^{1}$
Oleg V. Egorov,$^{1,5}$
\newauthor
Anastasia V. Kasparova,$^{1}$
Sergey A. Khoperskov,$^{3,4}$
Roman I. Uklein,$^{5}$
Olga V. Vozyakova$^{1}$
\\
$^1$ Sternberg Astronomical Institute, Moscow M.V. Lomonosov State University, Universitetskij pr., 13,  Moscow, 119234, Russia\\
$^2$ Smithsonian Astrophysical Observatory, 60 Garden Street MS09, Cambridge, MA 02138, USA\\
$^3$ Institute of Astronomy, Russian Academy of Sciences, Pyatnitskaya st., 48, 119017 Moscow, Russia\\
$^4$ GEPI, Observatoire de Paris, PSL Universit{\'e}, CNRS, 5 Place Jules Janssen, 92190 Meudon,
France\\
$^5$ Special Astrophysical Observatory, Russian Academy of Sciences, Nizhniy Arkhyz, Karachai-Cherkessian Republic, 357147, Russia\\
}

\begin{document}
\label{firstpage}
\pagerange{\pageref{firstpage}--\pageref{lastpage}} \pubyear{2018}
\maketitle

\begin{abstract}
The formation scenario for giant low surface brightness (gLSB) galaxies with discs as large as 100 kpc still remains unclear. These stellar systems are rare and very hard to observe, therefore a detailed insight on every additional object helps to understand their nature. Here we present a detailed observational study of the gLSB UGC~1922 performed using deep optical imaging and spectroscopic observations combined with archival ultraviolet data. We derived spatially resolved properties of stellar and ionized gas kinematics and characteristics of stellar populations and interstellar medium. We reveal the presence of a kinematically decoupled central component, which counter rotates with respect to the main disc of UGC~1922. The radial metallicity gradient of the ionised gas is in agreement with that found for moderate-size LSB galaxies. At the same time, a slowly rotating and dynamically hot central region of the galaxy hosts a large number of old metal-rich stars, which creates an appearance of a giant elliptical galaxy, that grew an enormous star forming disc. We reproduce most of the observed features of UGC~1922 in $N$-body/hydrodynamical simulations of an in-plane merger of giant \emph{Sa} and \emph{Sd} galaxies. We also discuss alternative formation scenarios of this unusual system.
\end{abstract}

\section{Introduction}\label{intro}

The processes of formation and evolution of disc galaxies are still not fully understood. In particular, there exists a class of objects whose formation is difficult to explain in the framework of a standard theory of galaxy evolution. These challenging systems are giant low surface brightness galaxies (gLSBG). They have enormous discs reaching 250~kpc in diameter \citep{Boissier2016} with the B-band central surface brightness fainter than 22~mag~arcsec$^{-2}$. It is difficult to form such discs in the hierarchical clustering paradigm where dark haloes of disc galaxies do not undergo major mergers. The prototype of this galaxy class is Malin~1 discovered by \citet{Bothun1987}. Deep photometric observations revealed very extended faint spiral structure that surrounds early-type galaxy with high surface brightness (HSB) disc and bulge \citep[see][]{Boissier2016,Galaz2015}. Several other examples of gLSBGs were discussed e.g. in \citet{Pickering1997}. The sample is being extended as new objects are discovered using improved observational facilities. For example, \citet{Hagen2016} found that UGC~1382, a system that was previously thought to be an elliptical galaxy, is, in fact a HSB lenticular galaxy with an extended gLSB disc similar to Malin~1. This study also raises a question on how well we understand the evolution of early type galaxies. How many ``normal'' elliptical galaxies at first sight are not what is expected and harbor giant LSB discs? This question remains open. So is the question of the formation of gLSBG. Several formation scenarios were proposed. \citet{Mapelli2008} proposed that gLSBG could be the results of a bygone head-on collision of the galaxy with a massive intruder. The ring caused by the collision expands and forms the giant disc in their models. \citet{Reshetnikov2010} considered this scenario as possible for Malin~1. However, a more recent study by \citet{Boissier2016} pointed out that the morphology and colors of the disc of Malin~1 contradict to the catastrophic scenario by \citet{Mapelli2008}. Also, \citet{Hagen2016} demonstrated that a ring galaxy was not an ancestor of UGC~1382. Instead, they considered another widely discussed way of the formation of gLSBGs proposed by \citet{Penarrubia2006}, where the gLSB disc is formed by accretion and tidal disruption of small gas-rich satellite galaxies. \citet{Hagen2016} found possible satellite remnant in the LSB disc of UGC~1382. However, the model by \citet{Penarrubia2006} predicts the decrease of the rotational velocity at the periphery of a disc, while the \HI rotation curves of several gLSBG remain flat to the outermost measured point \citep{Mishra2017}.  \citet{Kasparova2014} did not find traces of recent mergers in the colour maps of the gLSBG Malin~2. This suggests that the satellite tidal disruption scenario of the gLSBG formation is at least not the only one to be considered.

Another catastrophic scenario was proposed recently by \citet{Zhu2018} who found Malin~1 analogue in IllustrisTNG simulations and traced back its formation history. They found out that the gas rich giant discy galaxy appeared as a result of merger with two massive intruders with the peak circular velocities of 350 and 120 \kms. The mechanism proposed by \citet{Zhu2018} has the advantage that it can reproduce within  current
galaxy formation theory the extended disc together with flat rotation curve out to 200 kpc and spirals and clumps in the disc which are observed in gLSBs.  

Some authors propose non-catastrophic scenarios of the formation of gLSBGs. \citet{Noguchi2001} discussed the transformation of normal HSB spirals to gLSBGs through dynamical evolution due to the bar, which induces non-circular motions and radial mixing of disc matter that flattens the disc density profile. This model however has its own weak points -- most of gLSBg do not possess strong bars. \citet{Kasparova2014} proposed another non-catastrophic solution. They concluded that the gLSB disc of Malin~2 could have been formed because of unusual properties of the dark halo of the galaxy, namely, its large radial scale and low central density. \citet{Saburova2018} came to a similar conclusion for a sample of giant disc HSB galaxies.

In agreement with previous studies, \citet{Hagen2016} noticed that low density environment could be crucial for the preservation and possibly formation of giant discs. \citet{Saburova2018} also concluded that giant HSB discy galaxies tend to be slightly more rare in groups and in clusters rather than in the field.

To make further progress in understanding of the evolution and formation history of gLSBGs, we studied in detail one of the Malin~1-class galaxy, UGC~1922.   We give the basic properties of the system in Table \ref{properties}. It is a spiral galaxy with a prominent bulge embedded in the disc with asymmetric spiral arms featuring ``rows'' and irregular cloudy structure on the NW-side (see the $g$-band image in Fig.~\ref{map}). The position of this structure coincides with the direction to the galaxy LEDA~1829911 which does not have a spectroscopic redshift. If we assume they belong to the same group, the projected distance between the two galaxies will be about 120~kpc. According to \citet{Saulder2016}, UGC~1922 belongs to a group that includes 7 members (based on 2MRS data). In the current paper we adopt the distance of 150~Mpc for UGC~1922 as in \citet{Mishra2017}.  It corresponds to the scale of $0.73$ kpc/arcsec. The giant elliptical galaxy IC~227 with the optical radius of 43~kpc has the radial velocity different by $\sim$600~\kms\, from UGC~1922 and is located about 200~kpc away from it in projection. Thus, UGC~1922 is definitely not an isolated system. Therefore, we cannot exclude that irregular features of its disc and spiral arms can be a result of interaction with neighbouring galaxies.

Because of the low surface brightness of the disc, UGC~1922 was erroneously classified as an elliptical galaxy in the past  \citep[see][]{Huchra2012}, which makes it similar to UGC~1382 mentioned above.  By the size of its disc, the fraction of \HI and the value of dynamical mass UGC~1922 is similar to Malin~2.

The northern part of spiral structure and the central region are visible in archival GALEX NUV and FUV images \citep{Martinetal2005} which can indicate the presence of active star formation in these regions. The UV-brightness of the centre of UGC~1922 is slightly enhanced. The Galactic extinction for the galaxy is $E(B-V)=0.118^m$ according to \citet{Schlafly2011}.
\begin{table}
\caption{Basic properties of UGC~1922. \newline 
References: {[1]} NED (http://ned.ipac.caltech.edu), {[2]} \citet{Mishra2017}. \label{properties}
\label{tab1}}
 \begin{center}
 \begin{tabular}{lll}
\hline \hline
Names& UGC~1922&ref.\\
        & IC0226&\\
        & PGC009373&\\
\hline
Equatorial coordinates  & 02h27m45.9s &\\ 
(J2000.0)& +28d12m32s&[1]\\
Distance     & $150$ Mpc& [2]\\
Morphological type     & S?&[1]\\
Inclination angle     & $51$\degr&[2]\\
Major axis position angle    &$128$\degr&[2]\\
\hline
\end{tabular}

\end{center}
\end{table}

\citet{Schombert1998} found signs of an active galactic nucleus in the optical spectrum of UGC~1922. \citet{Mishra2017} observed the galaxy in \HI and found the central deficiency in \HI distribution which can also be an indirect indication of an AGN. The \HI density map traces the NUV-bright spiral arm but the \HI velocity map appears to be undisturbed.

We preformed long-slit spectral observations and deep multi-colour photometry. It allowed us to obtain the kinematical profiles of stars and ionized gas and also properties of stellar population and the gas phase metallicity, which we used in an attempt to build a more complete picture of this unusual system.

The current paper is organized as follows: Section \ref{Obs} is devoted to the details of observation and data reduction; the results of the data analysis are given in Section \ref{Res}; our attempt to determine the masses of the disc, bulge and dark matter halo is described in Section \ref{massmod}; we present discussion together with the results of $N$-body simulations aimed to reproduce the general features of the galaxy in Section \ref{Discussion}; the main results are summarized in Section \ref{conclusion}.

\section{Observations and data reduction}\label{Obs}
\subsection{Long-slit Spectroscopic Observations}
\begin{figure}
\vspace{-7cm}
\hspace{-0.7cm}
\includegraphics[width=\linewidth]{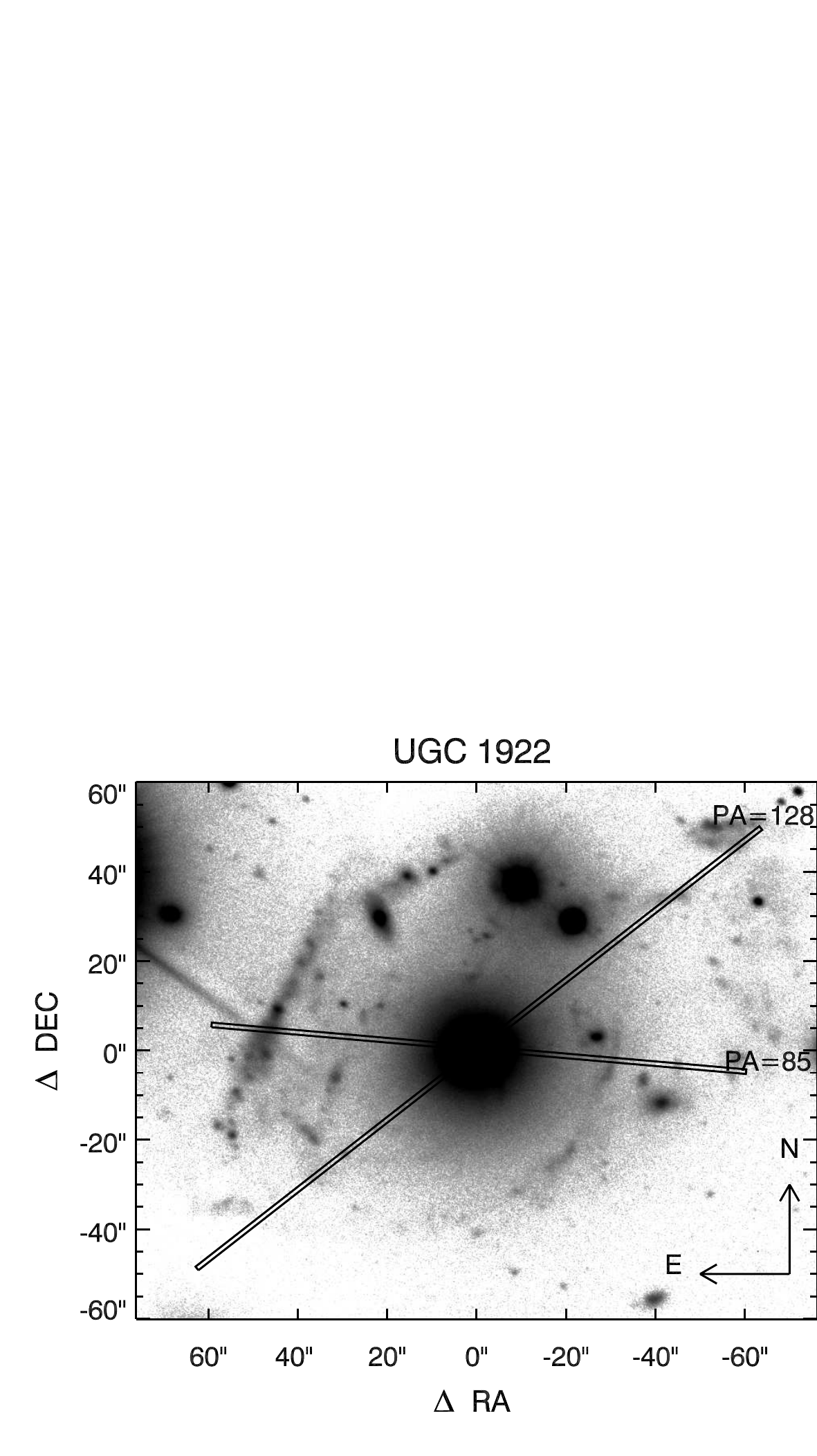}
\caption{The positions of the slits used during spectroscopic observations with the Russian 6-m BTA telescope overplotted on the $g$-band image obtained with 2.5-m telescope of the Sternberg Astronomical Institute.}
\label{map}
\end{figure}

We performed long-slit spectral observations of UGC~1922 with the spectrographs SCORPIO-2 \citep{AfanasievMoiseev2011} and SCORPIO \citep{AfanasievMoiseev2005} operated at the prime focus of the 6-m Russian telescope BTA at Special
Astrophysical Observatory of the Russian Academy of Sciences (SAO RAS). We show the positions of the slit on the g-band image obtained with the 2.5-m telescope of the Sternberg Astronomical Institute in Fig. \ref{map}. The dates of observations, exposure times, atmospheric seeing quality and the dispersers are listed in Table \ref{logsp}. The dispersers listed in Table \ref{logsp} have the following parameters. The grism VPHG1200@540 covers the spectral range 3600--7070 \AA\, and has a dispersion of 0.87 \AA\ pixel$^{-1}$, the instrumental FWHM of $\approx 5.2$ \AA\,  and the grism VPHG2300G possesses  the spectral range of 4800-5570 \AA, dispersion 0.38 \AA\ pixel$^{-1}$ and the instrumental FWHM of 2.2 \AA. The scale along the slit is 0.36~arcsec pixel$^{-1}$, the slit width was 1~arcsec.

\begin{table}
\caption{Log of spectral observations}\label{logsp}
\begin{center}
\begin{tabular}{ccccc}
\hline\hline
Slit PA & Date & Exp. time & Seeing& Disperser \\
   (\degr)  &  &     (s) &        (\arcsec) & \\
\hline
128&22.09.2017&7200&1.5 & VPHG2300G (Sco1) \\
85&25.11.2017&15600&1.8 & VPHG1200@540 (Sco2) \\
\hline
\end{tabular}
\end{center}
\end{table}

We performed the spectral data reduction using the \textsc{idl} based pipeline.  The reduction included the following steps: bias subtraction and truncation of overscan regions, flat-field correction, the wavelength calibration based on the spectrum of He-Ne-Ar  lamp\footnote{To achieve more accurate wavelength solution we calibrated separately the spectra with total exposure times 7200 s.}, cosmic ray hit removal, linearization, summation, the night sky subtraction using the algorithm described in \citet{KatkovChilingarian2011}  and flux calibration using spectrophotometric stellar standards BD33d2642 and BD25d4655.

We derived the parameters of the instrumental line spread function of the spectrographs from fitting of the twilight sky spectrum observed in the same observation runs. We fitted the reduced spectra of UGC~1922 with high-resolution PEGASE.HR~\citep{LeBorgneetal2004} simple stellar population models (SSP)  for Salpeter IMF \citep{Salpeter1955} convolved with the resulting instrumental profile. We performed this fitting using the \nb{} full spectral fitting technique  \citep{Chilingarian2007a, Chilingarian2007b}, which  allows to fit the spectrum in a pixel space. In \nb{} the parameters of the stellar populations are derived by nonlinear minimization of the quadratic difference chi-square between the observed and model spectra.  The non-linear minimization is performed using Levenberg-Marquardt minimization implemented in the {\sc mpfit} {\sc IDL} package  \citep[by C. Markwardt, NASA,][]{mpfit}. The parameters of SSP that we utilized are the age T~(Gyr) and metallicity [Fe/H]~(dex) of stellar population\footnote{ The ages and metallicities in the model  can have values: $T=$ 0.03, 0.05, 0.1, 0.2, 0.3, 0.5, 0.8, 1, 1.5, 2, 3, 4, 5, 6, 7, 8, 9, 10, 11, 12, 13, 14, 16, 18, 20 Gyr  and $Z=$-2.5, -2.0, -1.5, -1.0, -0.5, -0.3, 0.0, 0.3, 0.5, 1.0 dex.  }. The line-of-sight velocity distribution (LOSVD) of stars were parametrized by Gauss-Hermite series  \citep[see][]{vanderMarel1993}.  As the result of fitting we obtained the luminosity-weighted stellar age and metallicity, line-of-sight velocity, velocity dispersion and Gauss-Hermite moments $h_3$ and $h_4$ which characterize the deviation of LOSVD from the Gaussian profile. We defined the bins for the fitting manually.

We calculated the  uncertainties of the parameters of our stellar population model using  Monte Carlo simulations for a hundred realizations of synthetic spectra for each spatial bin which were created by adding a random noise corresponding to the signal-to-noise ratio in the bin to the best-fitting model.

Along with absorption we considered also the emission spectra which we got by subtraction of model stellar spectra from the observed ones. After that we fitted the emission lines in each pixel along the slit by Gaussian profiles and obtained as the result the velocity and velocity dispersion of ionized gas and fluxes in the emission lines. In addition we have measured the emission lines fluxes in the stacked spectra of several observed clumps or the areas of extended ionized gas along the slit PA=85\degr. The obtained results for each region are shown in Table~\ref{tab:fitres} including the position along the slit; the shift along RA and DEC from the galaxy centre; the reddening corrected fluxes in each measured emission line with their uncertainties; the adopted colour excess E(B-V)\footnote{For a few regions we obtained E(B-V) < 0. This could be due to uncertain subtraction of the underlying Balmer lines from stellar population; we adopted the value E(B-V)=0 for them.} measured by Balmer decrement made using \cite{Cardelli1989} extinction curve; and the measured oxygen abundance 12+$\log$(O/H).\footnote{Excluding several regions with high contribution of non-photoionization mechanism of emission lines excitation using several methods (see Section~\ref{sec:metall}).}

\begin{table*}
\begin{scriptsize}
\caption{\textbf{Measured fluxes in emission lines and oxygen abundance of the observed ionized clumps or extended regions of diffuse ionized gas together with the positions of the regions along the slit and reddening.}}\label{tab:fitres}
\begin{tabular}{lccccccc}
\hline
Region & 1 & 2 & 3 & 4 & 5 & 6 & 7 \\
\hline
Pos. along slit, arcsec & $-61.4 \div -57.8$ & $-48.6 \div -45.7$ & $-33.6 \div -30.0$ & $-29.6 \div -27.5$ & $-26.8 \div -23.9$ & $-23.2 \div -14.6$ & $-12.9 \div -9.3$ \\
$\Delta$RA, arcsec & $-40.2$ & $-28.0$ & $-13.0$ & $-9.8$ & $-6.7$ & $-0.4$ & $7.3$ \\
$\Delta$DEC, arcsec & $8.8$ & $7.0$ & $4.7$ & $4.2$ & $3.7$ & $2.8$ & $1.6$ \\
$F_\mathrm{H\beta}, 10^{-17}\ \mathrm{erg\ cm^{-2}\ s^{-1}\ arcsec^{-2}}$ &$ 1.22 \pm 0.24$ &$ 1.45 \pm 0.51$ &$ 3.61 \pm 0.40$ &$ 2.23 \pm 1.39$ &$ 1.25 \pm 0.32$ &$ 2.07 \pm 0.51$ &$ 2.37 \pm 0.37$ \\
{}[O \textsc{ii}] 3727,29\AA & $-$  & $-$  & $-$  & $-$  & $-$  & $-$  & $-$  \\
{}[O \textsc{iii}] 5007\AA & $ 1.39 \pm 0.16 $  & $ 2.29 \pm 0.29 $  & $ 1.23 \pm 0.09 $  & $ 1.90 \pm 0.50 $  & $ 0.47 \pm 0.33 $  & $ 0.65 \pm 0.22 $  & $ 0.63 \pm 0.10 $  \\
{}[O \textsc{i}] 6300\AA & $-$  & $-$  & $-$  & $-$  & $-$  & $-$  & $-$  \\
{}[O \textsc{i}] 6364\AA & $-$  & $-$  & $-$  & $-$  & $-$  & $-$  & $-$  \\
{}[N \textsc{ii}] 6548\AA & $ 0.33 \pm 0.09 $  & $ 0.46 \pm 0.14 $  & $ 0.35 \pm 0.04 $  & $ 0.38 \pm 0.14 $  & $ 0.37 \pm 0.11 $  & $ 0.60 \pm 0.12 $  & $ 0.42 \pm 0.07 $  \\
H$\alpha$ & $ 2.87 \pm 0.12 $  & $ 2.91 \pm 0.21 $  & $ 2.89 \pm 0.07 $  & $ 2.94 \pm 0.19 $  & $ 2.88 \pm 0.17 $  & $ 2.88 \pm 0.23 $  & $ 2.87 \pm 0.10 $  \\
{}[N \textsc{ii}] 6584\AA & $ 1.01 \pm 0.11 $  & $ 1.39 \pm 0.20 $  & $ 1.06 \pm 0.06 $  & $ 1.14 \pm 0.19 $  & $ 1.11 \pm 0.15 $  & $ 1.84 \pm 0.21 $  & $ 1.29 \pm 0.10 $  \\
{}[S \textsc{ii}] 6717\AA & $ 0.80 \pm 0.18 $  & $ 1.01 \pm 0.27 $  & $ 0.65 \pm 0.07 $  & $ 0.81 \pm 0.22 $  & $ 0.85 \pm 0.20 $  & $ 1.38 \pm 0.25 $  & $ 1.14 \pm 0.12 $  \\
{}[S \textsc{ii}] 6731\AA & $ 0.47 \pm 0.16 $  & $ 0.82 \pm 0.31 $  & $ 0.40 \pm 0.08 $  & $ 0.51 \pm 0.23 $  & $ 1.02 \pm 0.30 $  & $ 0.87 \pm 0.27 $  & $ 0.83 \pm 0.14 $  \\
E(B-V), mag & $ 0.05 \pm 0.17 $  & $ 0.38 \pm 0.31 $  & $ 0.21 \pm 0.10 $  & $ 0.65 \pm 0.54 $  & $ 0.17 \pm 0.23 $  & $ 0.12 \pm 0.22 $  & $ 0.09 \pm 0.14 $  \\
12+$\log$(O/H)$_\mathrm{S}$ & $8.50 \pm 0.10$ & $-$ & $8.54 \pm 0.05$ & $-$ & $8.45 \pm 0.14$ & $-$ & $8.54 \pm 0.07$ \\
12+$\log$(O/H)$_\mathrm{O3N2}$ & $8.41 \pm 0.05$ & $-$ & $8.42 \pm 0.03$ & $-$ & $8.51 \pm 0.10$ & $-$ & $8.51 \pm 0.05$ \\
12+$\log$(O/H)$_\mathrm{izi}$ & $8.54 \pm 0.13$ & $-$ & $8.57 \pm 0.08$ & $-$ & $8.60 \pm 0.20$ & $-$ & $8.63 \pm 0.10$ \\
12+$\log$(O/H)$_\mathrm{HCm}$ & $8.57 \pm 0.07$ & $-$ & $8.55 \pm 0.07$ & $-$ & $8.70 \pm 0.09$ & $-$ & $8.70 \pm 0.07$ \\
\hline
\hline
Region & 8 & 9 & 10 & 11 & 12 & 13 & 14 \\
\hline
Pos. along slit, arcsec & $-7.9 \div -4.3$ & $-2.9 \div 2.1$ & $3.6 \div 6.4$ & $7.5 \div 11.8$ & $26.4 \div 30.7$ & $43.2 \div 46.1$ & $47.1 \div 52.8$ \\
$\Delta$RA, arcsec & $12.2$ & $17.8$ & $23.1$ & $27.6$ & $46.1$ & $61.9$ & $67.1$ \\
$\Delta$DEC, arcsec & $0.9$ & $0.0$ & $-0.8$ & $-1.5$ & $-4.3$ & $-6.7$ & $-7.5$ \\
$F_\mathrm{H\beta}, 10^{-17}\ \mathrm{erg\ cm^{-2}\ s^{-1}\ arcsec^{-2}}$ &$ 4.19 \pm 0.33$ &$ 132.31 \pm 2.15$ &$ 5.98 \pm 0.39$ &$ 4.28 \pm 0.38$ &$ 3.01 \pm 0.31$ &$ 1.43 \pm 0.34$ &$ 3.29 \pm 0.44$ \\
{}[O \textsc{ii}] 3727,29\AA & $-$  & $ 5.23 \pm 0.36 $  & $-$  & $-$  & $-$  & $-$  & $-$  \\
{}[O \textsc{iii}] 5007\AA & $ 0.85 \pm 0.07 $  & $ 0.95 \pm 0.02 $  & $ 0.33 \pm 0.08 $  & $ 0.31 \pm 0.06 $  & $ 0.63 \pm 0.08 $  & $ 1.02 \pm 0.19 $  & $ 0.75 \pm 0.13 $  \\
{}[O \textsc{i}] 6300\AA & $-$  & $ 0.66 \pm 0.01 $  & $ 0.53 \pm 0.05 $  & $-$  & $-$  & $-$  & $-$  \\
{}[O \textsc{i}] 6364\AA & $-$  & $ 0.16 \pm 0.01 $  & $ 0.02 \pm 0.04 $  & $-$  & $-$  & $-$  & $-$  \\
{}[N \textsc{ii}] 6548\AA & $ 0.67 \pm 0.05 $  & $ 0.78 \pm 0.01 $  & $ 0.51 \pm 0.04 $  & $ 0.27 \pm 0.05 $  & $ 0.36 \pm 0.05 $  & $ 0.33 \pm 0.11 $  & $ 0.32 \pm 0.06 $  \\
H$\alpha$ & $ 2.86 \pm 0.07 $  & $ 2.91 \pm 0.03 $  & $ 2.62 \pm 0.05 $  & $ 1.54 \pm 0.07 $  & $ 2.87 \pm 0.10 $  & $ 2.88 \pm 0.17 $  & $ 2.88 \pm 0.11 $  \\
{}[N \textsc{ii}] 6584\AA & $ 2.04 \pm 0.07 $  & $ 2.37 \pm 0.03 $  & $ 1.56 \pm 0.05 $  & $ 0.83 \pm 0.07 $  & $ 1.11 \pm 0.09 $  & $ 1.00 \pm 0.17 $  & $ 0.97 \pm 0.10 $  \\
{}[S \textsc{ii}] 6717\AA & $ 1.33 \pm 0.07 $  & $ 1.54 \pm 0.02 $  & $ 0.93 \pm 0.06 $  & $ 0.43 \pm 0.06 $  & $ 0.70 \pm 0.08 $  & $ 0.92 \pm 0.16 $  & $ 0.77 \pm 0.10 $  \\
{}[S \textsc{ii}] 6731\AA & $ 1.03 \pm 0.08 $  & $ 1.25 \pm 0.02 $  & $ 0.79 \pm 0.07 $  & $ 0.41 \pm 0.08 $  & $ 0.51 \pm 0.10 $  & $ 0.49 \pm 0.18 $  & $ 0.62 \pm 0.13 $  \\
E(B-V), mag & $ 0.03 \pm 0.07 $  & $ 0.36 \pm 0.02 $  & $-$  & $-$  & $ 0.07 \pm 0.10 $  & $ 0.17 \pm 0.21 $  & $ 0.13 \pm 0.12 $  \\
12+$\log$(O/H)$_\mathrm{S}$ & $-$ & $-$ & $8.60 \pm 0.03$ & $8.58 \pm 0.09$ & $8.53 \pm 0.06$ & $8.50 \pm 0.15$ & $8.47 \pm 0.09$ \\
12+$\log$(O/H)$_\mathrm{O3N2}$ & $-$ & $-$ & $8.60 \pm 0.03$ & $8.57 \pm 0.05$ & $8.48 \pm 0.04$ & $8.41 \pm 0.08$ & $8.46 \pm 0.05$ \\
12+$\log$(O/H)$_\mathrm{izi}$ & $-$ & $-$ & $8.70 \pm 0.07$ & $8.63 \pm 0.13$ & $8.60 \pm 0.09$ & $8.57 \pm 0.17$ & $8.54 \pm 0.11$ \\
12+$\log$(O/H)$_\mathrm{HCm}$ & $-$ & $-$ & $8.75 \pm 0.05$ & $8.72 \pm 0.12$ & $8.70 \pm 0.06$ & $8.61 \pm 0.09$ & $8.65 \pm 0.07$ \\
\hline
\end{tabular}
\end{scriptsize}
\end{table*}

\subsection{Imaging Data}
We obtained deep {\it g,r}-band images of UGC~1922 with 2.5-m telescope of the Sternberg Astronomical Institute. We give the details of observations in Table \ref{logph}. The primary reduction was performed using the pipeline written in Python and included overscan and non-linearity corrections and division by normalized flat-field. After that we did the images alignment, cosmic ray hit correction, summation and sky subtraction by \textsc{idl} based pipeline. We made the photometric calibration using the aperture photometry of stars visible in the images
with the magnitudes available in the Data Release 1 of the Pan-STARRS1 Surveys \citep[PS1,][]{Chambers2016_panstarrs1}. We obtained the following transformation equations:
\begin{align}
m_g=35.433\pm0.07 -2.5\log(F);\\
m_r=34.890\pm0.08 -2.5\log(F)
\end{align}
where F is instrumental flux.
\begin{table}
\caption{Observational log for imaging}\label{logph}
\begin{center}
\begin{tabular}{ccccc}
\hline\hline
Band & Date & Exposure time& Seeing& Observer \\
   (\degr)  &  &     (s) &        (\arcsec) & \\
\hline
{\it g}&25.10.2017&1800&1.2 & Tatarnikov \\
{\it r}&25.10.2017&1500&1.2 & Tatarnikov \\
\hline
\end{tabular}
\end{center}
\end{table}

\section{The results of observations}\label{Res}
\subsection{Radial profiles of kinematics and the properties of stellar population}

\begin{figure*}
\includegraphics[width=0.5\linewidth]{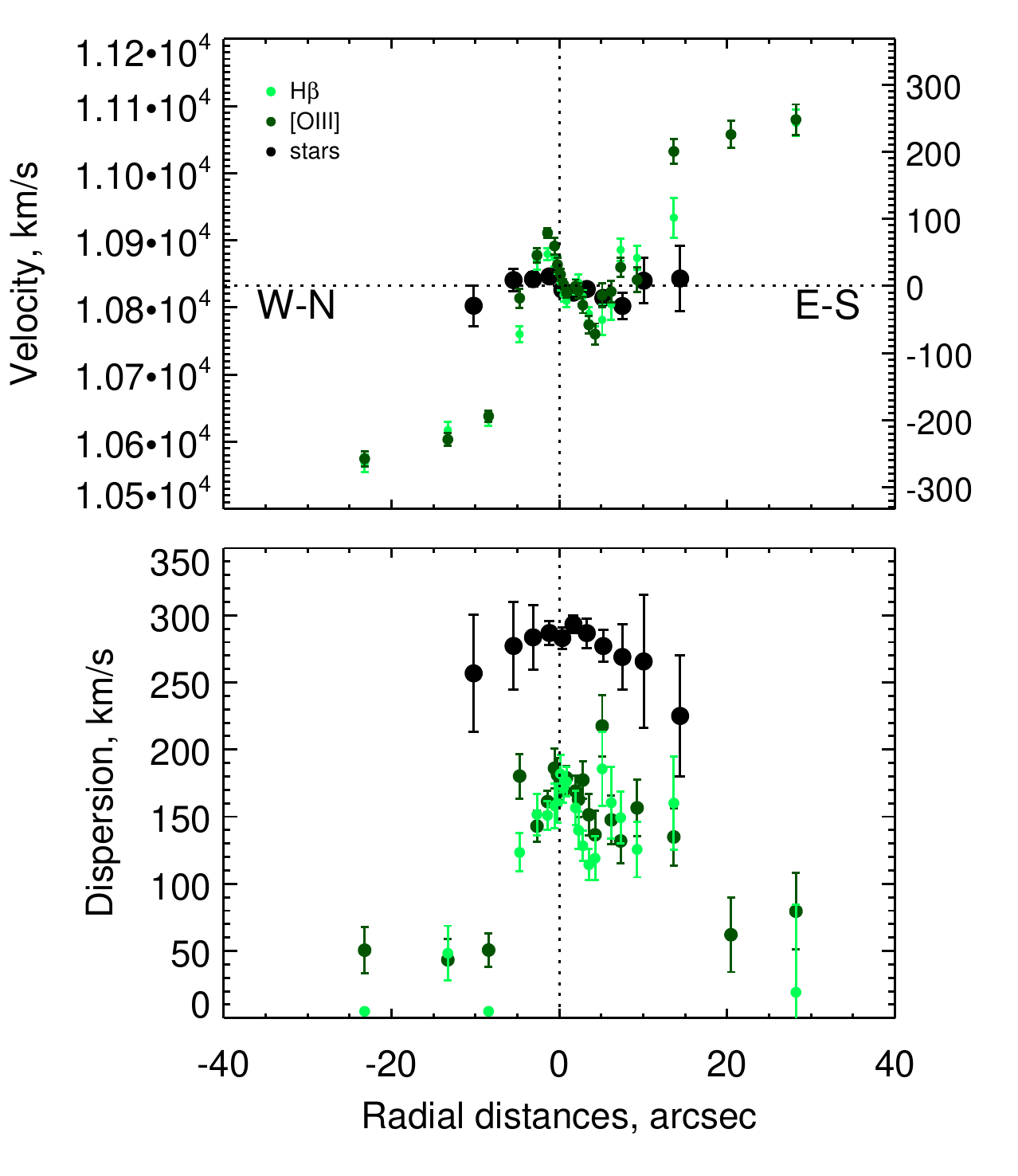}
\includegraphics[width=0.5\linewidth]{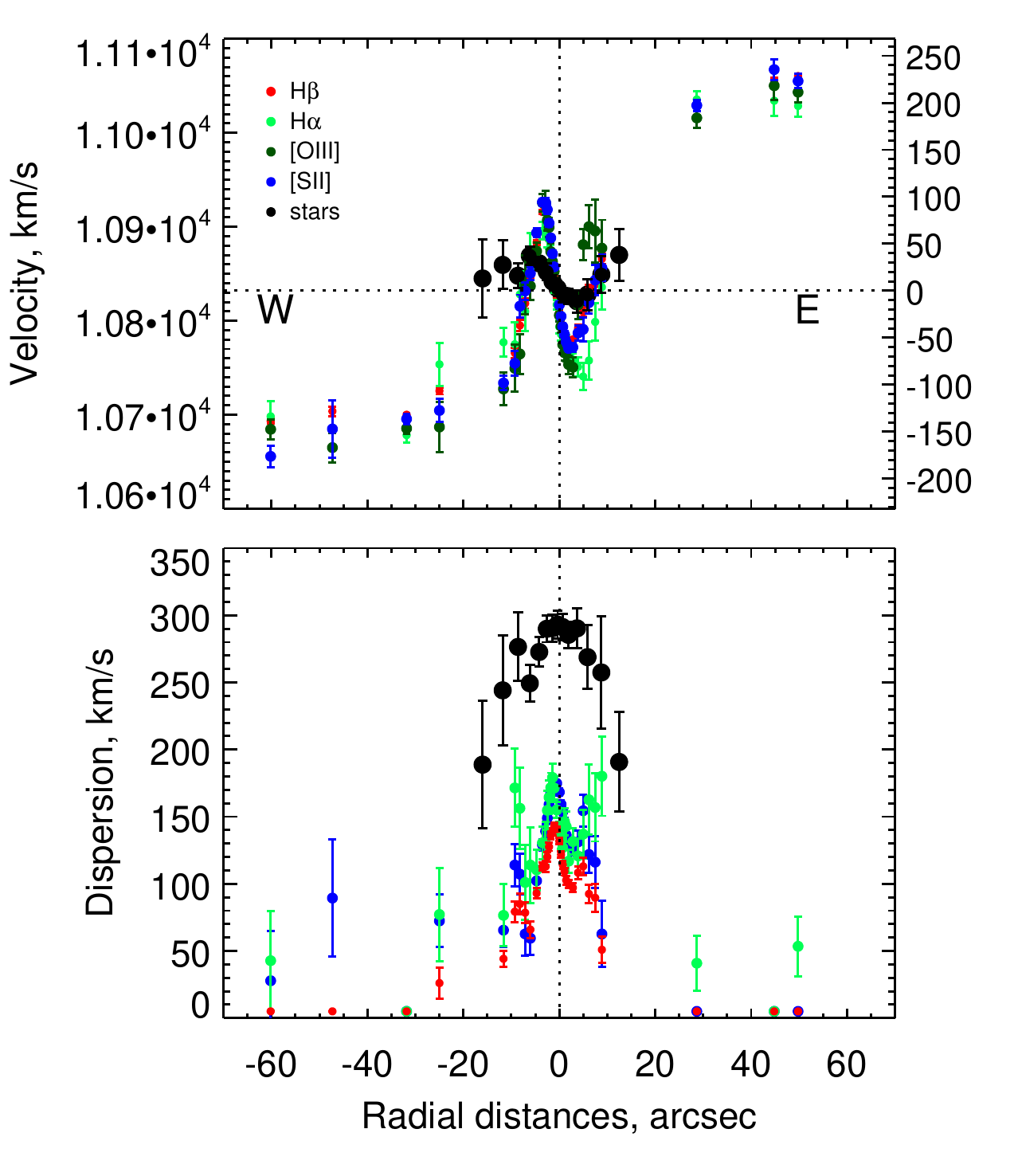}
\caption{The profiles of the line-of-sight velocity (top panel) and velocity dispersion (bottom panel) for PA=128\degr\ (left column) and PA=85\degr\ (right column). Coloured symbols show the ionised gas kinematics in different optical emission lines, black circles show stellar kinematics.}
\label{kin_profiles}
\end{figure*}

In Fig. \ref{kin_profiles} we demonstrate the profiles of line-of-sight (LOS) velocity and velocity dispersion of ionized gas and stars for PA=128\degr\ (left-hand column) and PA=85\degr\ (right-hand column). The values of $h_3$ and $h_4$ are within the range of $-0.1...0.1$. The most prominent feature of the profiles of LOS velocity is the presence of counter-rotation of gas in the central region with respect to the outer part. The variation of the velocity of stars also follows the behavior of ionized gas kinematics (see the right-hand column of Fig. \ref{kin_profiles}), although the amplitude of the change of the velocity for stars is lower than that of the gas. The velocity dispersion of stars is high in the centre  ($\sim 300$~\kms) and shows only slight hint to central minimum which could be expected if there was a nuclear disc. 

\begin{figure*}

\includegraphics[width=0.5\linewidth]{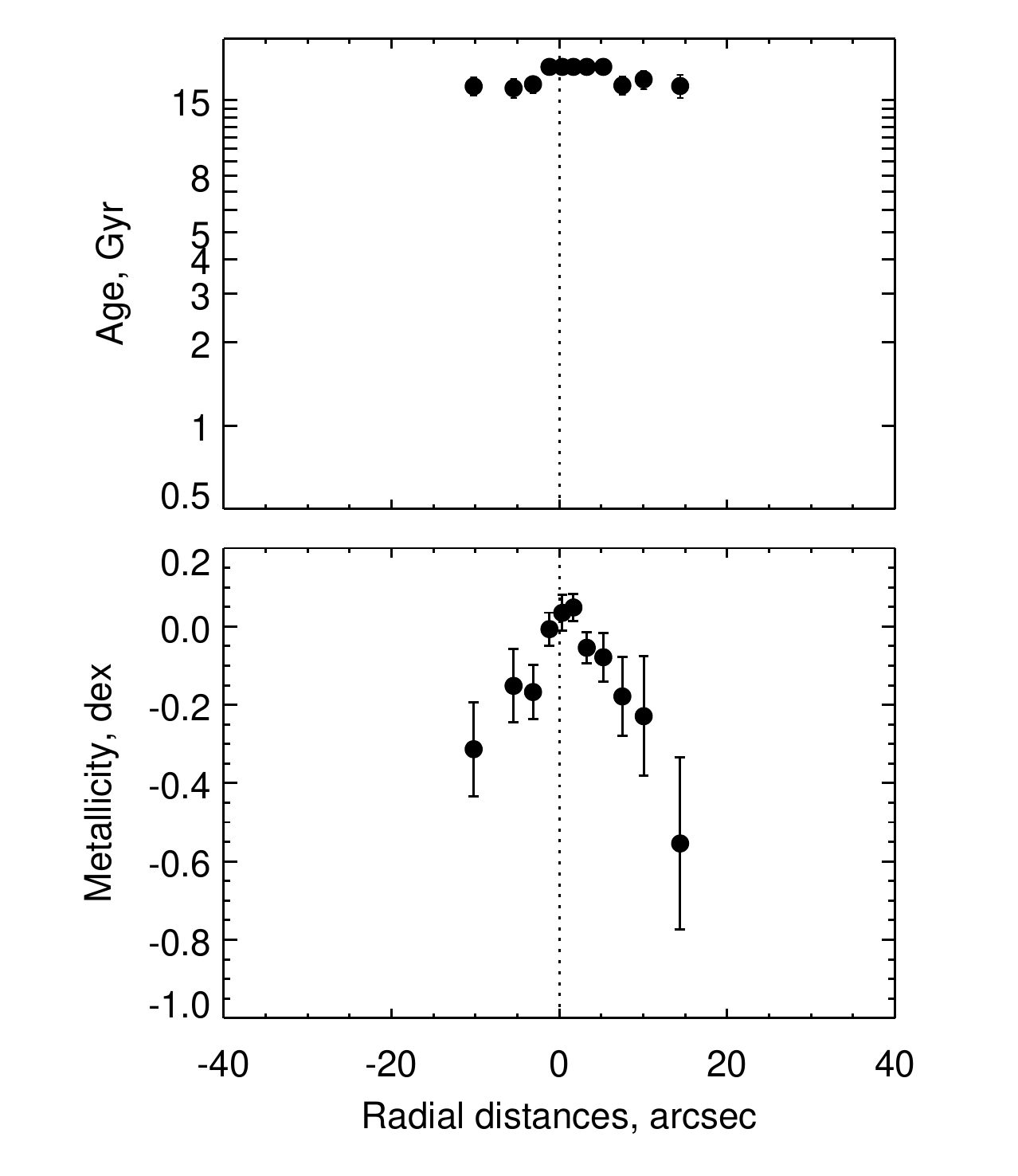}
\includegraphics[width=0.5\linewidth]{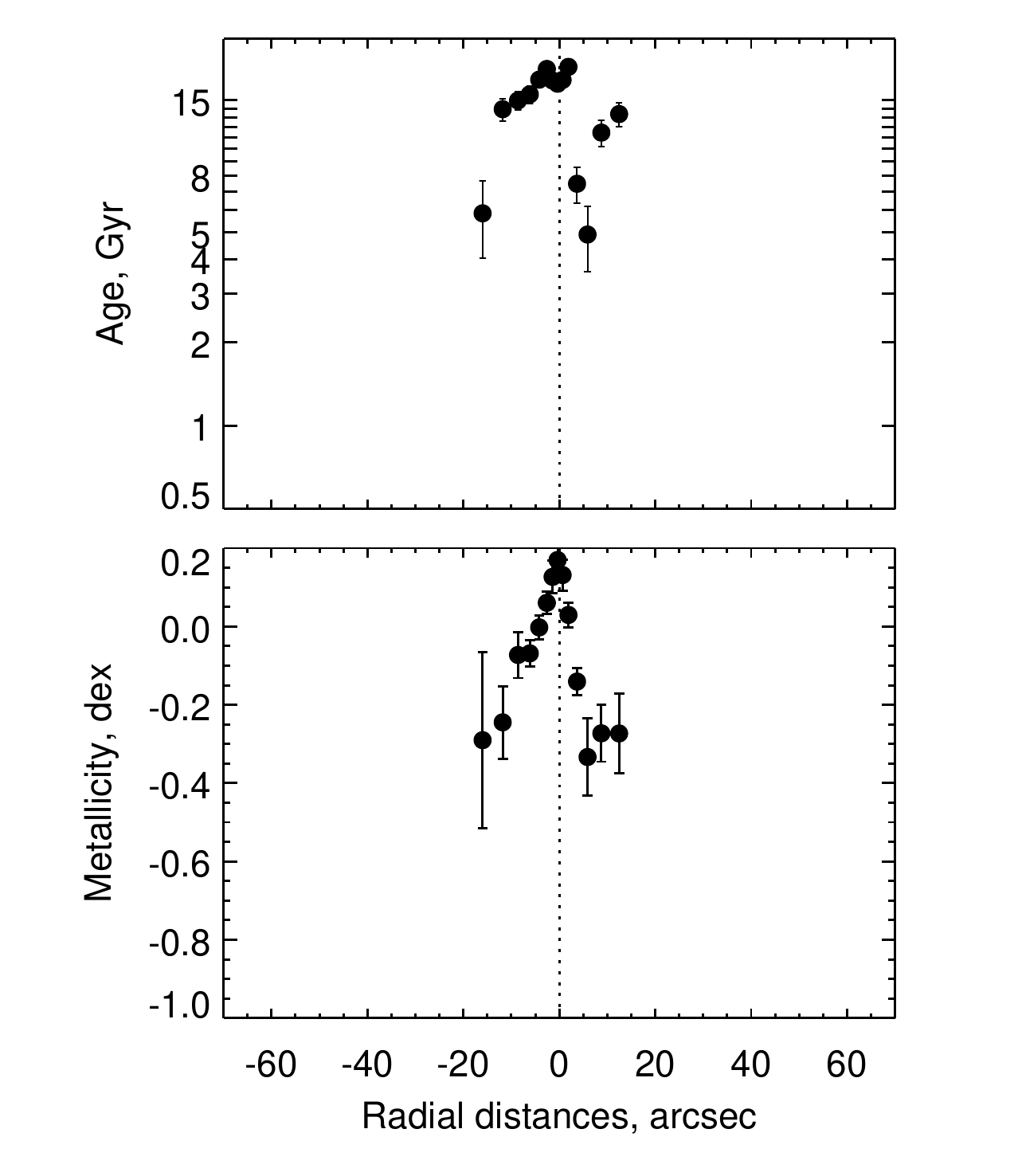}
\caption{The radial profiles of SSP equivalent age (top panels) and metallicity (bottom panels) of stars for PA=128\degr\ (left panels) and PA=85\degr\ (right panels). }

\label{pop_profiles}

\end{figure*}

One can see that the age of stellar population is very old in the centre --  higher than 13.8 Gyr (see Fig. \ref{pop_profiles}, where we show the profiles of metallicity and age of stars for both position angles). The age of stars of LSB disc of UGC~1922 is lower than that of the central region (being, however, not younger than 5-6 Gyr), so is the stellar metallicity. The stars in the centre have almost solar metallicity. 

  One should keep in mind that SSP model that we used here may be over-simplified especially for the complex system like UGC~1922. Below we show that in the innermost region its both spectral energy distribution (SED) and spectrum could be better fitted by the model with two bursts of star formation. However, this more complex analysis is reasonable to apply only with SED, since it bears information in UV-range in which the young stars vastly manifest themselves, which is obstructed by the quality of the UV-data.

\subsection{The results of two-component-fitting of central spectrum and SED}\label{2c}
To understand what caused counter-rotation in the central region of UGC~1922 we decided to test if our data are compatible with the presence of two stellar components in the centre. One component could be related to the less-massive gas-rich galaxy which could be accreted by the bulge of UGC~1922 and is a possible origin of counter-rotation. Another component is old bulge. To do it we fitted the integrated spectrum of the central area with the radius of 1 arcsec for PA=128\degr\ and the spectral energy distribution (SED) for central 6 arcsec (the resolution was limited by the PSF of GALEX data) using the {\sc nbursts+phot} technique described in detail in \citet{ChilingarianKatkov2012}. This technique is extension of the {\sc nburst} full spectrum fitting which allows one to include photometric
constraints into the fit.   The algorithm
first fits a spectrum and obtains $\chi^2_{\mathrm{spec}}/D.O.F.$ and then
applies derived relative weights of stellar population components to compute
$\chi^2_{\mathrm{SED}}/D.O.F.$ from the broad-band photometry.

\begin{figure*}
\includegraphics[width=\linewidth]{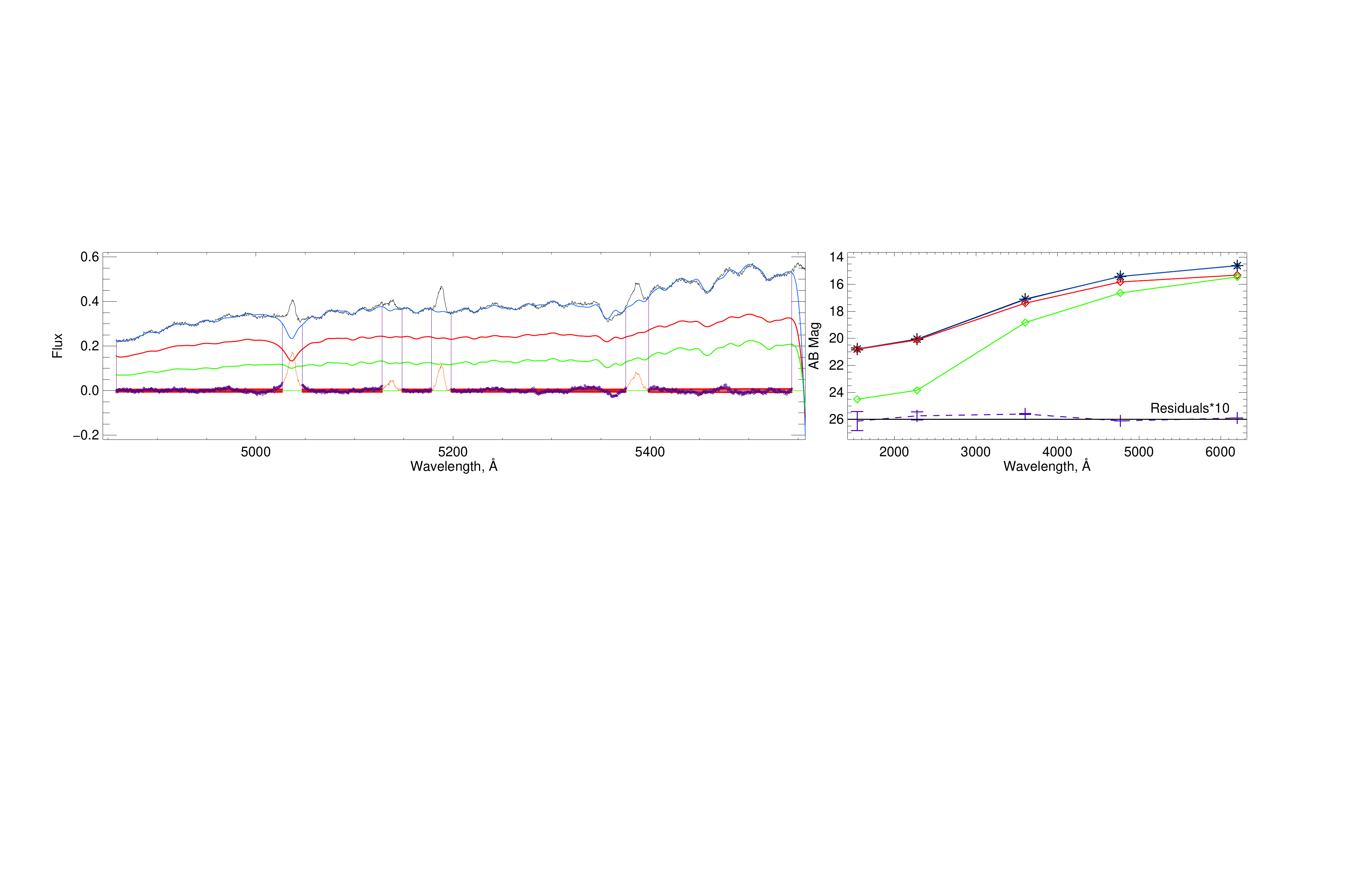}
\vspace{-5.5cm}
\caption{The result of spectral (left-hand panel) and SED (right-hand panel) fitting for central region of UGC~1922. Red and green lines show the contributions of two stellar components. Blue line denotes the total model. The fitting residuals are demonstrated in the bottom of each plot. Black symbols show the observed data. }
\label{2compfit}
\end{figure*}

To get SED we calculated the FUV, NUV, $u$, $g$, $r$-band magnitudes  of the central region of the galaxy. The $u$-band image was taken from the Canada-France-Hawaii Telescope (CFHT) data-archive, $g$, $r$-images were obtained in current work. To take into account the PSF of GALEX images we convolved $u$, $g$, $r$-band images with Gaussian with FWHM = 6 arcsec. We took into account both for internal E(B-V)=0.37 (calculated from Balmer ratio $H_\alpha / H_\beta$) and Galactic extinction. 
We show the results of the fitting in Fig. \ref{2compfit}. As one can see the observed data can be satisfactorily explained by the two stellar components  ($(\chi^2)_{red}=1.943$).  The fits for single component had significantly worse values of the reduced $\chi^2$ for both simple SSP model (3.015) and model with exponential star formation history (2.858).  The two components have the following ages and metallicities:  $T_1= 0.4$ Gyr and  $T_2$  older than 13.8 Gyr,  $Z_1=-0.07$ dex and  $ Z_2=  -0.29$ dex. Thus, our data agree with the presence of young stellar component which could be related to the fossil of the gas-rich  galaxy which fell into the bulge of UGC~1922 and caused counter-rotation. The second old component is the bulge. 

\subsection{The metallicity and excitation mechanism of ionised gas}\label{sec:metall}
\begin{figure*}
\includegraphics[width=\linewidth]{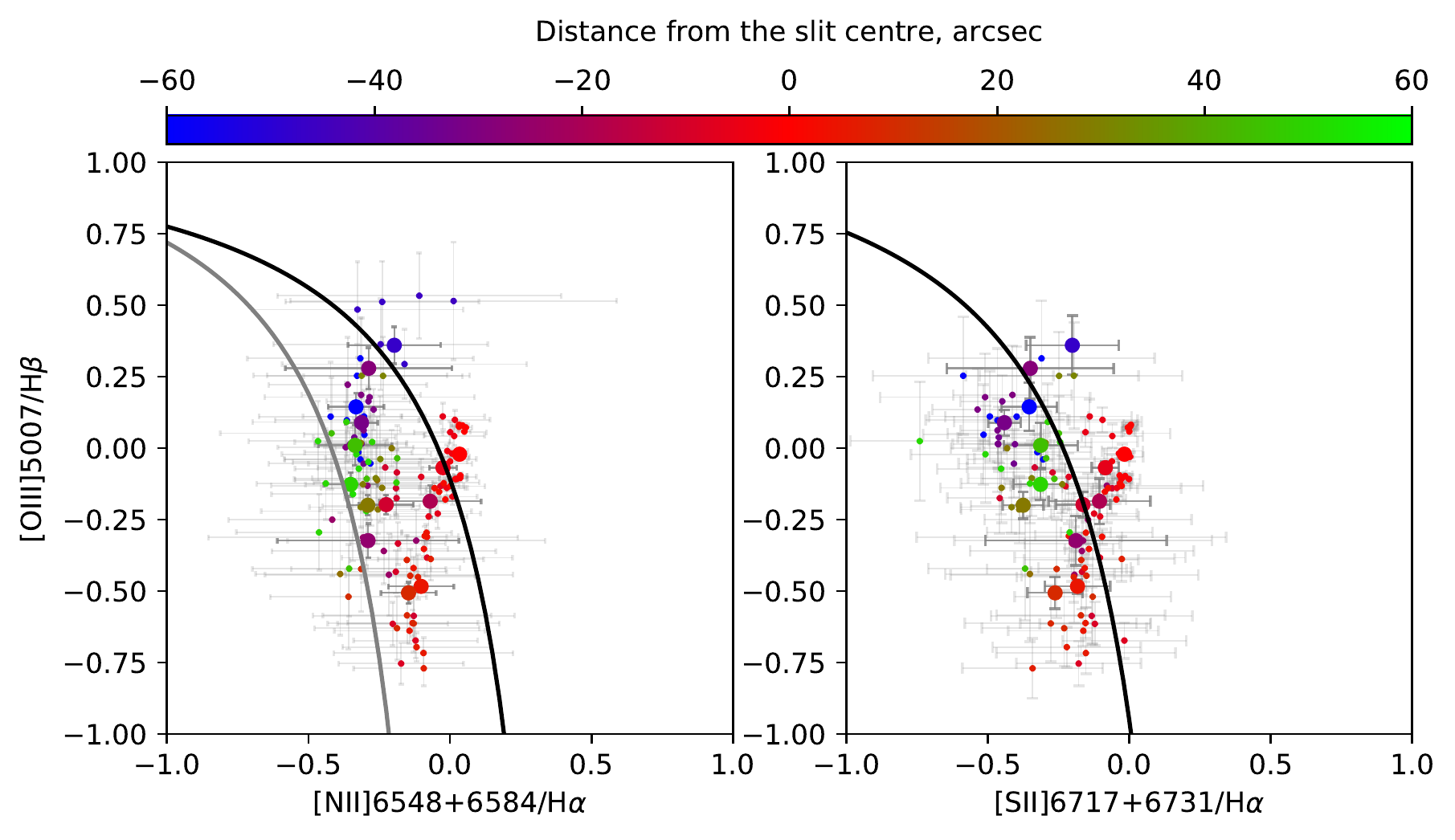}
\caption{The BPT diagrams for UGC~1922 (PA=85\degr). The colour shows the distance from the centre. Large symbols correspond to the integrated spectra of separate clumps,  for which we give coordinates along the slit in Table \ref{tab:fitres}, small circles demonstrate the position of each pixel along the slit.The grey line taken from  \citep{Kewley2001} separates photoionised \HII-regions and all other types of gas excitation; black line from \citet{Kauffmann03} separates the regions with composite mechanism of excitation.}
\label{bpt}
\end{figure*}

From the spectral data obtained with SCORPIO-2 (PA=85\degr) with the wide wavelength coverage, we can conclude on the processes responsible for excitation of the ionised gas in the bulge and the disc of UGC~1922. In Fig. \ref{bpt} we show emission line ratios on the BPT diagnostics diagram \citet{BPT}. The colour of the data points indicates the radial distance from the centre. Large symbols correspond to the integrated spectra of separate clumps  (see Table \ref{tab:fitres} for their positions), while small circles demonstrate the position of each pixel along the slit.  From the diagram one can see that most of the regions of the disc lie between two separation lines on the left panel and hence exhibit the composite mechanism of the emission lines excitation. Both photoionization by massive stars and non-photoionization mechanisms should play role there. In the central part of the galaxy, the gas is ionized by a harder radiation or by a collisional mechanism. It can also indicate weak nuclear activity in this system. One region of the disc also lies above the demarcation line and corresponds to hard excitation mechanism. This region is a part of the irregular clumpy structure in the NW side of the disc which could be the trace of interaction. It can indicate the presence of the strong non-circular or slowly decaying high velocity turbulent motions that may result in the higher input of collisional excitation into the gas emission in this area. The opposite side of the disc lies closer to the region of the photoionization excitation.

Another noticeable feature of the BPT-diagram is its bimodality. The central region is shifted along both x- and y-axis with respect to the disc region.  This result is in good agreement with the finding by \citet{CALIFA2015} for a sample of 306 galaxies who concluded that regions situated in the central areas of galaxies, and in earlier type galaxies are located in lower-right side of the distribution on BPT-diagram, while those regions situated at larger galactocentric distances and in later type galaxies are more frequently located towards the upper-left side of the distribution. According to \citet{CALIFA2015} the lower-right side of the BPT-diagram also corresponds to higher ages and metallicities of the stellar population and higher oxygen abundance. The metallicity and the age of stellar population are indeed lower in the disc of UGC~1922 in comparison to the bulge region, as well as the gas oxygen abundance.

\begin{figure}
\includegraphics[width=\linewidth]{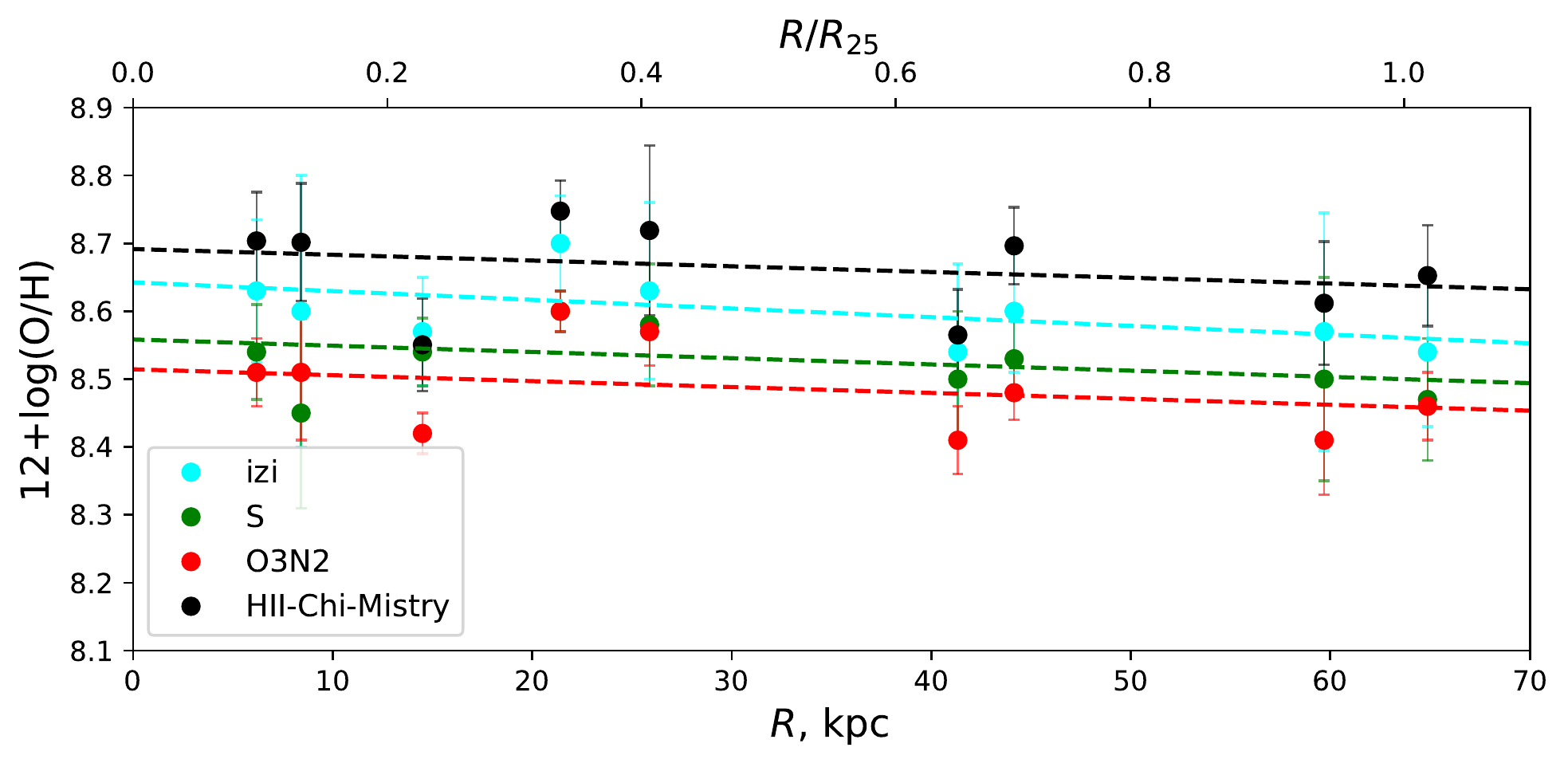}
\caption{The oxygen abundance $12+\log(\mathrm{O/H})$ radial profile plotted for UGC~1922 PA=85\degr. The deprojected radii are expressed in terms of kpc (bottom x-axis scale) and of  $R_{25}\approx 4(R_d)_r\approx64$ kpc (top x-axis scale). Different colours denote the corresponding methods for estimating the oxygen abundance.}
\label{oh}
\end{figure}

A precise estimate of the oxygen abundance, which is a measure of the gas phase metallicity, is very complex because of well known and still unresolved problems of the inconsistency between estimates (and even its gradient) derived with different methods \citep[see, e.g.,][]{Kewley2008, Lopez-Sanchez2012}. We were not able to use a so-called ``direct'' $T_e$-based method because of the weakness of the lines sensitive to the electron temperature; also we were unable to use any method based on [O~{\sc ii}] line because of the poor quality of the blue-end of the spectrum. Here we use several empirical and model-based calibrators; they are denoted further as O3N2 \citep{Marino2013}, S \citep{Pilyugin16}, HII-Chi-Mistry \citep{Perez-Montero2014}, and \textsc{izi} \citep{Blanc2015} with \citet{Levesque2010} photoionization models. 
All these methods use the ratio of the fluxes of forbidden lines [O~{\sc iii}], [N~{\sc ii}] (and some of them -- [S~{\sc ii}]) to recombination H$\alpha$ or H$\beta$ lines. The results of their application to our data are shown in Fig.~\ref{oh}, where all points sitting above the ``maximum starburst line'' on BPT diagram are excluded from the analysis.

As it follows from the oxygen abundance analysis, all methods used in our study reveal a shallow metallicity radial gradient ($0.06-0.09 \pm 0.10$ dex/$R_{25}$, or $0.03-0.04 \pm 0.05$ dex/$R_{d}$).
These values are only slightly lower than the metallicity gradients for disc galaxies when normalizing to their $R_e$ (or $R_d$): $\sim 0.07$ dex/$R_{d}$ by CALIFA data \citep{Sanchez2014} and $\sim 0.05$ dex/$R_{d}$ by MANGA data \citep{Belfiore2017}. \citet{Ho2015} obtained a median metallicity gradient $\sim 0.39$ dex/$R_{25}$ for a large sample of local star-forming galaxies, that is much higher than our estimate. Hence, we may conclude that the metallicity gradient for UGC~1922 is shallower in comparison with normal galaxies, but comparable when normalized to its disc scale.

By analyzing spectra of 10 LSB galaxies of  moderate sizes, \citet{Bresolin2015} have obtained  similar result -- a measurable, but shallow gradient. Our derived values of the metallicity gradient for UGC~1922 agree well with the gradients found in that study  despite these systems could be of different nature since they are characterized by later morphological types and have significantly lower sizes of the discs. The authors showed that the metallicity gradients in both HSB and LSB galaxies are inversely proportional to their disc scale $R_{d}$, that could reflect the underlying physical mechanism like mass surface density -- metallicity relation (see their Fig.~3). Our estimates also clearly follow this trend. Therefore, no  {\it recent} episodes of the metal-poor gas accretion to the centre of the galaxy are needed to explain the observed shallow metallicity gradient in UGC~1922: it might be a result of a normal evolution in framework of the ``inside-out'' disc formation scenario.

\subsection{Radial profiles of surface brightness}

\begin{figure}

\includegraphics[width=\columnwidth,trim=0.3cm 0 0.3cm 0,clip]{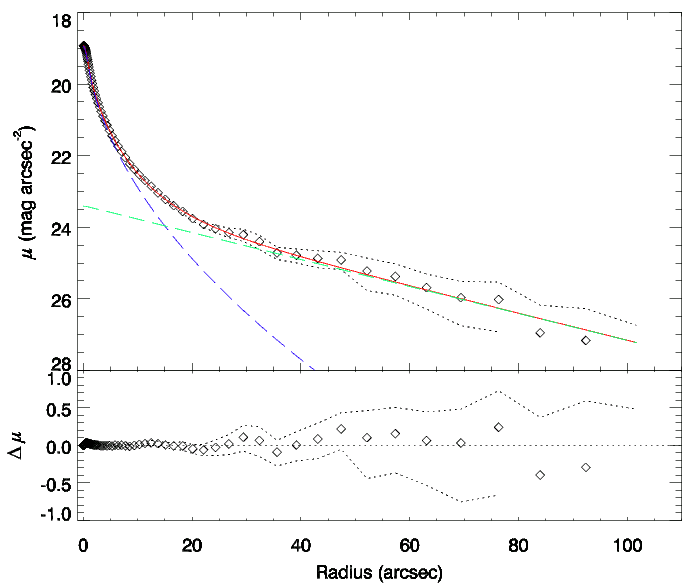}
\caption{Top panel: the {\it g}-band radial surface brightness profile uncorrected for extinction (diamonds) decomposed into the contributions of bulge (blue dashed line) and disc (green dashed line). Red thin line shows the model profile. Bottom panel demonstrates the residuals.}
\label{phot_profile}

\end{figure}

Using our deep {\it g,r}-band images, we derived the surface brightness profiles using {\sc ellipse} routine \citep{Jedrzejewski1987} in the {\sc iraf} software (\citealt{iraf}).
After that we decomposed the profiles into the contribution of exponential disc:
\begin{equation}I_d(r)=(I_{d})_{0}\exp(-r/R_d), \end{equation} where $(I_{d})_{0}$ and $R_d$ are the disc central surface brightness and the exponential scalelength, correspondingly and Sersic bulge \citep{Sersic68}:
\begin{equation}
\label{eq:Sersic}
I_b(r) =
          (I_0)_b 10^{\left[ - b_n\left( \frac{r}{R_{e}} \right)^{1/n} \right]}. \end{equation}
Here $(I_0)_b$ is the bulge central surface brightness, $R_e$ is the effective radius containing a half of the luminosity,
$b_n\approx 1.9992 n - 0.3271$ (\citealt{Caon1993})
and $n$ is the Sersic index.  To do it we used the \textsc{idl}-based code which utilizes Levenberg-Marquardt least-squares fit.  The resulting profile and its model is shown in Fig. \ref{phot_profile} for {\it g}-band. We give the  parameters of the disc and bulge in {\it g} and {\it r}-bands in Table \ref{phot_par}. The disc and bulge central surface brightnesses are corrected for Galactic extinction.

As one can see from Fig. \ref{phot_profile} and Table \ref{phot_par}, the radial surface brightness profiles of UGC~1922 are well described by the combination of a bulge and an LSB disc with no need for introducing the second HSB disc as for Malin~1  \citep[see][]{Lelli2010}. The Sersic index of bulge, however, is relatively low. The {\it g}-band disc radial scalelength is 21 kpc which corresponds to the disc radius $4R_d=84$ kpc - this size is comparable to that of Malin~1 and Malin~2.

\begin{table*}
\caption{The photometrical parameters of disc and bulge of UGC~1922: central surface brightness of bulge; effective radius of bulge; Sersic index of bulge; central surface brightness of exponential disc; radial scalelength of disc }\label{phot_par}
\begin{center}
\begin{tabular}{cccccc}
\hline\hline
Band&$(\mu_0)_b$&$R_e$&n&$(\mu_0)_d$&$R_d$\\
&(mag arcsec$^{-2}$)&(arcsec)&&(mag arcsec$^{-2}$)&(arcsec)\\
\hline
{\it g} &$ 17.78 \pm 0.01$ & $7.02 \pm 0.04$ & $1.97 \pm 0.02$ & $22.96 \pm 0.05$ & $28.8 \pm 1.0$\\
{\it r} & $16.94 \pm 0.02$ & $6.74 \pm 0.04$ & $1.98 \pm 0.03$ & $22.07 \pm 0.08$& $21.9 \pm 0.9$\\
\hline
\end{tabular}
\end{center}
\end{table*}

\section{The disc and dark halo masses of UGC~1922}\label{massmod}

\subsection{The disc mass estimate from the marginal gravitation stability criterion}
The mass contributions of the disc and dark halo can give us some clues on the nature of UGC~1922. The presence of counter-rotation complicates the modelling of the rotation curve of UGC~1922, so does the uncertain estimate of the inclination  \citep[see][]{Mishra2017}. However, we still can derive constrains on the masses of the disc and dark halo.

The independent upper disc mass estimate can be obtained from the stellar velocity dispersion profile and the assumption that the disc is marginally gravitationally stable  \citep[for more details see, e.g.][]{Zasov2004, Saburova2011, Saburova2013}. This assumption is confirmed by the majority of late-type disc galaxies and some of S0 galaxies \citep{Zasov2011} and could be formulated in the following way. A single component isothermal disc is locally marginally stable when the radial stellar velocity dispersion $c_r$ at the distance from the centre $R$ is equal to the critical value:

\begin{equation}
(c_{r})_{\rm crit}=Q_T\cdot 3.36G \sigma_d / \varkappa,
\end{equation}
where $\varkappa$ is the epicyclic frequency, and $Q_T$ is the Toomre's stability  parameter which is equal to unity for pure radial perturbations of a thin disc. Numerical simulations show that for the marginal stability of exponential discs with finite thickness, the parameter  $Q_T\approx 1.2 - 3$ is slowly growing with the radial distance \citep[e.g.][]{Khoperskov2003}.  The presence of cold gaseous component can make the disc more unstable \citep[se, e.g.][]{Romeo2011}, however the gaseous surface density in UGC~1922 is by an order lower than that of the stars at the considered radius, so we neglect this effect here. 

Thus, if we know the stellar radial velocity dispersion of the disc, we can calculate the disc surface density. We estimated the radial velocity dispersion from the observed line-of-sight stellar velocity dispersion $c_{\rm obs}$, taking into account the expected links between the dispersion along the radial, azimuthal and vertical directions:

\begin{multline}
c_{\rm obs}^2(r) = c_z^2 \cdot \cos^2 (i)+ c_{\phi}^2\cdot \sin^2 (i)\cdot \cos^2(\alpha) + \\ c_{r}\sin^2(i)\cdot \sin^2(\alpha),
\end{multline}
where $\alpha$  is the angle between the direction of the slit and the major axis.

To solve the equation we need two additional conditions: $ c_{r} = 2\Omega
\cdot  c_{\phi} /\varkappa $ (Lindblad formula for the epicyclic approximation) and $ c_z =k\cdot c_{r} $, where $c_z$, $c_{\phi}$, $c_{r}$ are the dispersion along the vertical, azimuthal and radial directions, projected on the plane of the galaxy. The coefficient $k$  was taken to be 0.6 in accordance with direct measurements, which show that it could lie in the range 0.5--0.8 \citep[see e.g.][]{Shapiro2003}. We derived the epicyclic frequency from the combined optical and \HI rotation curve \citep{Mishra2017} using the equation: $\varkappa(r)=2v(r)/r\sqrt{0.5+r/2v(r)( \frac{\partial v(r)}{\partial r})}$.

For the stellar velocity dispersion at the outermost reliable measured point ($R=16$~arcsec), we obtained the surface density of the disc of 270~\Msun pc$^{-2}$ which corresponds to the disc {\it r}-band mass-to-light ratio $M_d/L_r=11$ \Msun/ L\ensuremath{_\odot}.\footnote{We admit that this point is not in the region of a pure disc (i.e. with the negligible contribution of the bulge), which potentially makes our disc mass over-estimated. However, the estimate made for the outermost point $R=35$~arcsec ($c_{\rm obs}=140$~\kms) although characterized by high uncertainty corresponds to the close value of the disc surface density when interpolated to the same radius.} This value appears to be roughly five times as high as the one estimated from the corrected for Galactic extinction disc colour $(g-r)_0=0.72$ and model relations from \citet{Bell2003} and \citet{Roediger2015}: $M/L_r=2.44-3.1$ \Msun/ L\ensuremath{_\odot}.  This discrepancy could be even higher if the metallicity of the disc is low or if to take into account the possible variations of star formation history (e.g. for exponential star formation history).  It  either indicates the dynamical overheating of the disc -- the assumption of marginal gravitational stability is not valid for the disc of UGC~1922 (which seems more likely especially if to take into account the presence of possible traces of interaction in the disc), or the presence of large amount of unseen matter in the disc of UGC~1922 either in non-baryonic or baryonic form (i.e. cold gas non-detected by its emission, see, e.g. \citealt{Kasparova2014} or large amount of low-massive stars \citealt{Lee2004}). High dynamical mass-to-light ratios of LSB discs were previously noticed in other papers (see e.g. \citealt{Fuchs2003}; \citealt{Saburova2011}).

The way to check whether the disc of UGC~1922 could be significantly more massive than follows from the photometry is to test if such high contribution of the disc is compatible with observed rotation curve. To do it we made the combined rotation curve from our ionized gas data (the central region of the curve with counter-rotation was not considered here) and  \HI data from \citet{Mishra2017}. We adopted the position angle of the major axis and inclination given in Table \ref{tab1}. The inclination estimate according to \citet{Mishra2017}  is in good agreement with the results of our isophotal analysis.

\subsection{Mass modelling using the rotation curve}

\begin{figure*}

\includegraphics[trim=0.7cm 0 1cm 0,clip,height=0.22\textwidth ]{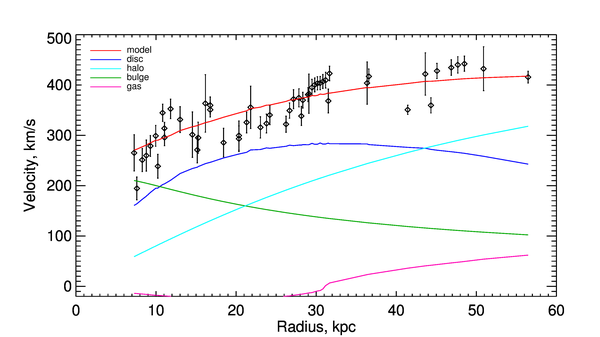}
\includegraphics[trim=2.55cm 0 1cm 0,clip,height=0.22\textwidth]{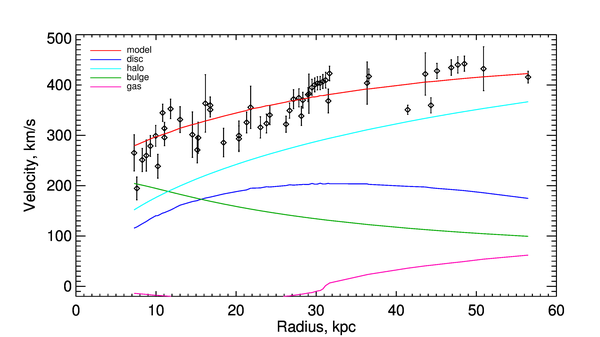}
\includegraphics[trim=2.55cm 0 1cm 0,clip,height=0.22\textwidth]{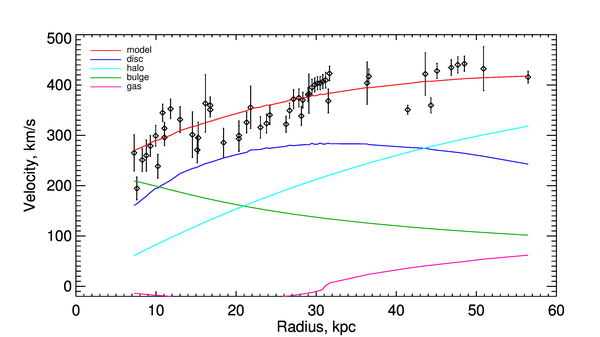}

\includegraphics[height=0.26\textwidth]{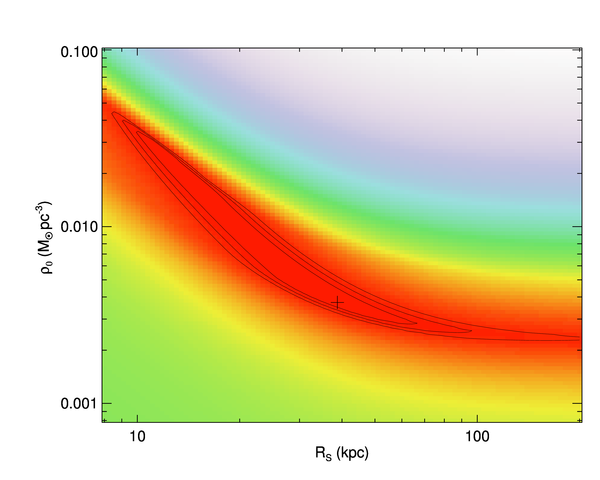}
\includegraphics[height=0.26\textwidth]{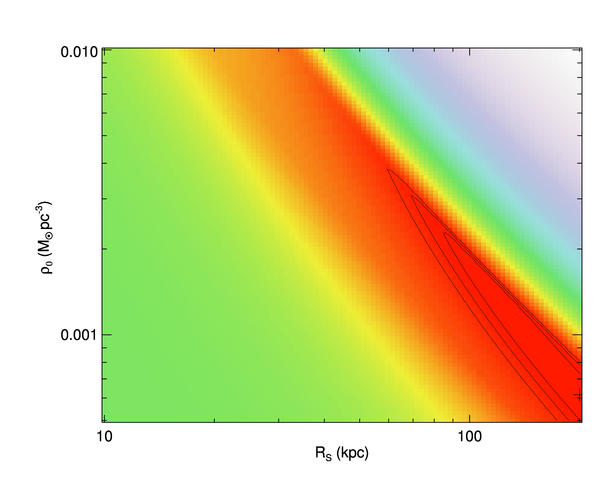}
\includegraphics[height=0.26\textwidth]{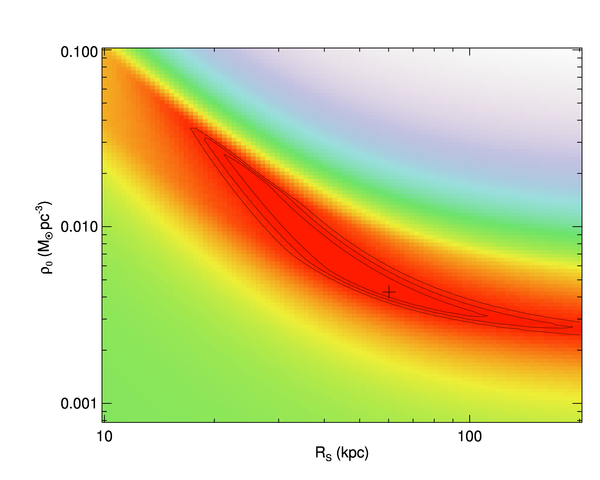}

\caption{Top panel: the best-fitting models of the combined \HI+ optical rotation curve (open symbols) left ~--- for the piso profile of the DM halo, centre ~--- for the NFW profile, right ~--- to the Burkert profile.
Bottom panel: $\chi^2$ map for the parameters of dark halo, color in the maps denotes the $\chi^2$ value, the darker the color, the lower the $\chi^2$ and the better is the fitting quality.
The contours refer to $1\sigma$, $2\sigma$ and $3\sigma$ confidence limits.
The position of the parameters corresponding to the $\chi^2$ minimum is shown by the cross in each map.}
\label{phot_mod}
\end{figure*}

We decomposed the combined rotation curve into the following components: a stellar exponential disc, a Sersic bulge and a dark matter halo and  gas disc\footnote{ We calculated the \HI surface density from the moment 0 maps from \citet{Mishra2017} using {\sc ellipse} routine in the {\sc iraf} software (\citealt{iraf}) and took into account the input of He.}. The discs and bulge rotation curves were constructed as described in \citet{Saburova2016}. We considered the following dark halo density profiles. (i) A density profile by \citet{Burkert1995}:
\begin{equation}\label{Burkert}
\rho_{\mathrm{burk}}(r)=\frac{\rho_0 R_s^3}{(r+R_s)(r^2+R_s^2)}.
\end{equation}
Here $\rho_{0}$ and $R_{s}$ are the central density and the radial scale of the halo\footnote{Below $R_s$ and $\rho_0$ are different for the various DM density profiles.}.
(ii) A pseudoisothermal profile (hereafter, piso):
\begin{equation}
\rho_{\mathrm{piso}}(r)=\frac{\rho_{0}}{(1+(r/R_{s})^2)},
\end{equation}
The Navarro-Frenk-White profile \citet{nfw1996}  (hereafter, NFW): \begin{equation}
\rho_{\mathrm{nfw}}(r)=\frac{\rho_{0}}{(r/R_s)(1+(r/R_{s })^2)^{2}},
\end{equation}
We fixed the disc and bulge radial scales from the {\it r}-band surface photometry. The mass-to-light ratio of the disc was restricted in the range $M/L_r=2.44...11$ \Msun/ L\ensuremath{_\odot} during the fitting (the first value comes from the photometry and the second follows from the marginal gravitational stability criterion). The mass-to-light ratio of the bulge was in the range: $2.4...5$ \Msun/ L\ensuremath{_\odot}.
In Fig. \ref{phot_mod} we show the results of the decomposition. The parameters of the components are given in Table \ref{par}. The best-fitting model appears to have close value of disc mass to what is expected from the marginal gravitational stability for  for piso and Burkert halo but not for NFW-halo. Thus such high disc mass-to-light ratio is not in conflict with observed rotation velocity amplitude.  However if one gives attention to the $1\sigma$ confidence limits of the disc mass-to-ligh ratio given in Table \ref{par}, one can see that the range of $M/L_r$ that follows from the best-fitting modelling includes the photometrically determined ratios. It indicates that the observed rotation curve does not allow to make choise between the heavy and light disc cases. Another concern is that due to the counter-rotating component we can not exclude the presence of non-circular motions that can make our rotation curve analysis less reliable.   

In order to have full information we also generated models with fixed mass-to-ligh ratios of the disc: maximum-disc model with  $M/L_r=11$ \Msun/ L\ensuremath{_\odot} and photometrical model with  $M/L_r=2.44$ \Msun/ L\ensuremath{_\odot}, the resulting parameters are given in Table \ref{par}.

\begin{figure*}
\includegraphics[trim=0 0 1cm 0,clip,width=\columnwidth]{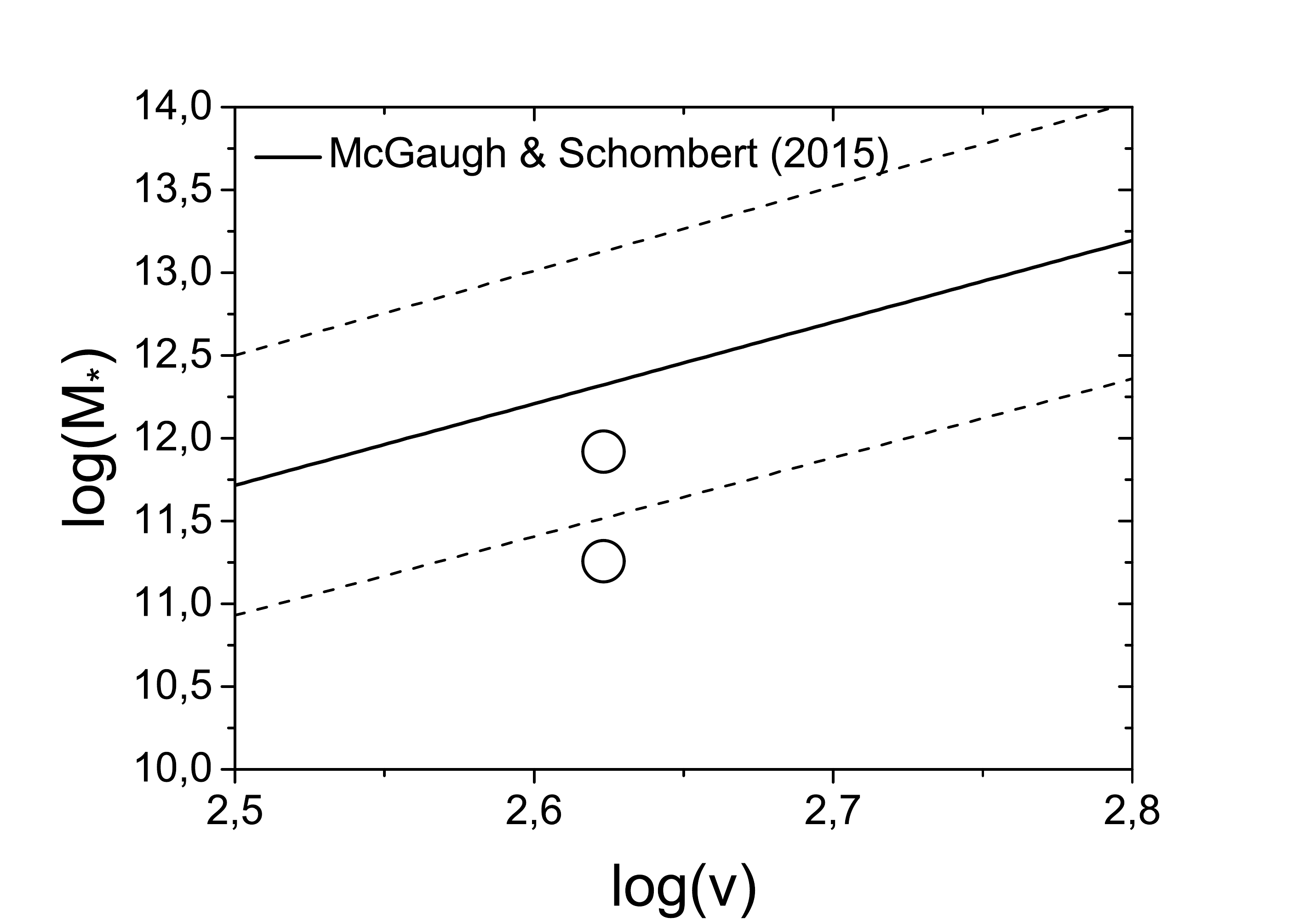}
\includegraphics[trim=0 0 1cm 0,clip,width=\columnwidth]{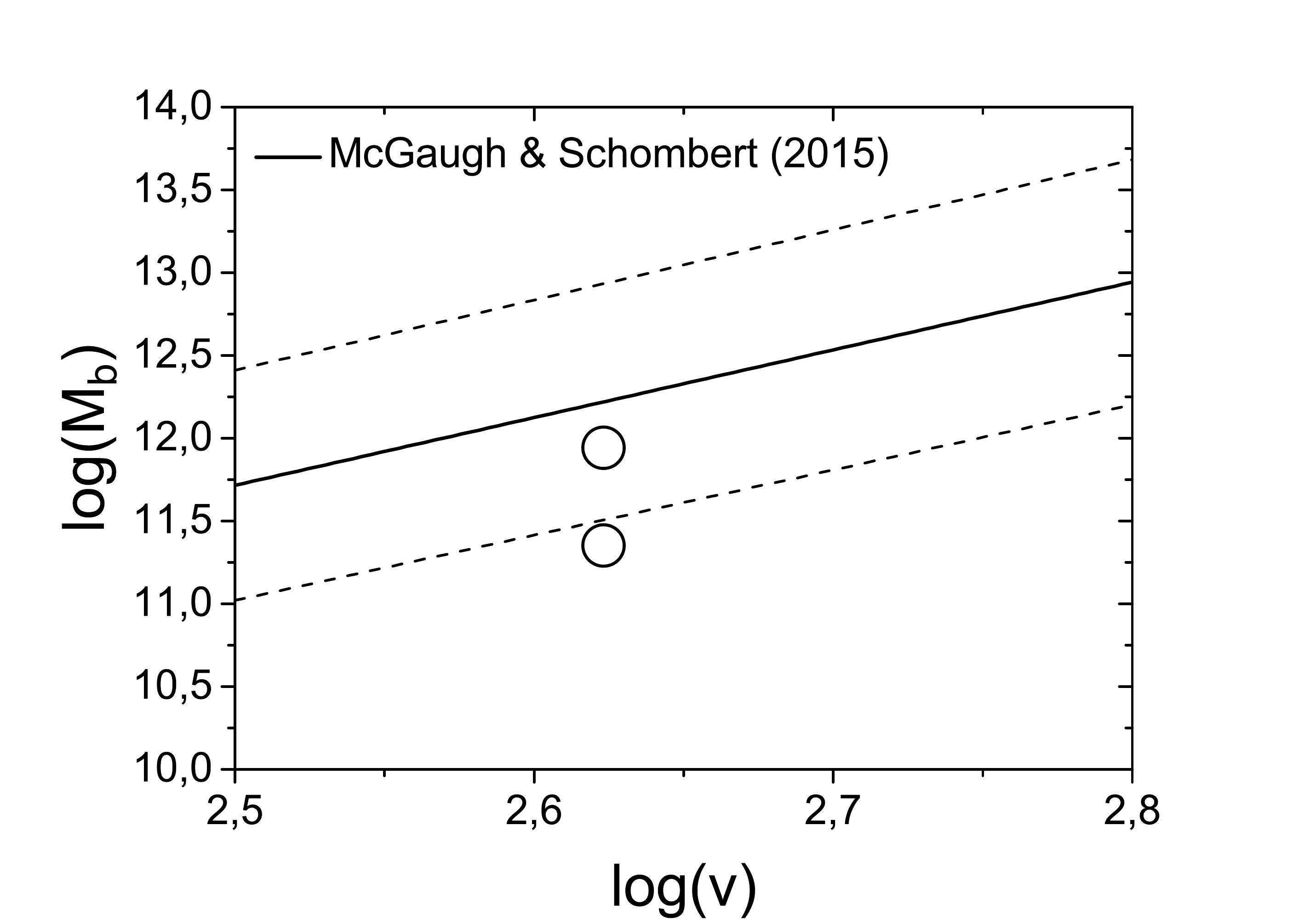}
\caption{The Tully-Fisher relation. Stellar mass vs. rotation velocity (left-hand column),  baryonic mass vs. rotation velocity (right-hand column).  The lines show the relations from \citet{McGaughSchombert2015}. Circles correspond to the disc mass-to-light ratios $M/L_r=11$ \Msun/ L\ensuremath{_\odot} following from the marginal gravitational stability criterion (upper circle) and the photometrical value $M/L_r=2.44$ \Msun/ L\ensuremath{_\odot} (lower circle).  The baryonic mass (right-hand panel) includes stellar mass and mass of gas from \citet{Mishra2017}.} 
\label{tf}
\end{figure*}

 To make a further check of the possibility of the heavy disc in UGC~1922, we also plotted both estimates of the mass of stellar disc (the photometric one and that following from the marginal stability condition) on the Tully-Fisher diagram where we compare the stellar mass and rotation velocity (see Fig. \ref{tf}). The line in Fig.~\ref{tf} corresponds to the relation found by \citet{McGaughSchombert2015}. We also compare the two masses with the baryonic Tully-Fisher relation found by the same authors. We derived the baryonic mass by adding the gas mass from \citet{Mishra2017} to the stellar one discussed above. One can see that both estimates do not strongly deviate from the relations found for other galaxies, however the mass calculated for marginally stable disc (upper point) is in better agreement with  both stellar and baryonic dependences from \citet{McGaughSchombert2015}. Thus, we can not exclude that the disc of UGC~1922 contains unseen matter. One should keep in mind, however that the marginal gravitational stability criterion gives  only the upper limit of stellar disc mass.
 
 Interestingly in the case of the heavy disc its mass makes roughly one third of the total mass within its radius, and is only two times lower than the mass of dark halo, which is typical for HSB galaxies  \citep[see, e.g.]{Zasov2011}. In case of the disc mass-to-light ratio taken from the photometry UGC~1922 is dark matter dominated system. However, in both cases we admit the high mass of the disc of UGC~1922 (see Table \ref{par}).

\begin{table*}
\begin{center}
\caption{The derived parameters of the main components of the galaxies.
The errors correspond to $1\sigma$ confidence limit. The columns contain the following data:
(1)~-- dark halo profile;
(2) and (3)~-- radial scale and central density of the DM halo;
(4)~-- optical radius defined from {\it r} band photometry;
(5)~-- mass of DM halo inside of optical radius;
(6)~-- disc {\it r} mass-to-light ratio;
(7)~-- central surface density of bulge
(8)~-- disc mass\label{par};}
\renewcommand{\arraystretch}{1.5}
\begin{tabular}{lrlrl r rlrlrlrc}
\hline
dark halo	&	\multicolumn{2}{c}{$R_s$}&	\multicolumn{2}{c}{$\rho_0$ }&	\multicolumn{1}{c}{$R_{\rm opt}$}	&	\multicolumn{2}{c}{$M_{\rm halo}$}	 &	\multicolumn{2}{c}{$M/L$} &	\multicolumn{2}{c}{$(I_0)_b$}&\multicolumn{1}{c}{$M_{\rm disc}$}\\
&\multicolumn{2}{c}{kpc}&\multicolumn{2}{c}{$10^{-3}$ M$_{\odot}/$pc$^3$}& \multicolumn{1}{c}{kpc}&\multicolumn{2}{c}{$10^{12}$ M$_{\odot}$}&\multicolumn{2}{c}{M$_{\odot}/$L$_{\odot}$	}& \multicolumn{2}{c}{$10^{3}$ M$_{\odot}/$pc$^2$}&\multicolumn{1}{c}{$10^{12}$ M$_{\odot}$}\\
\hline
\hline
\multicolumn{12}{c}{Best-fitting model }\\
\hline
          Burkert&      57.84 & $^{+     41.88}_{-     36.53} $  &      4.41& $^{+     21.11}_{-      1.28} $  &     63.70&    1.64& $^{+     0.59}_{-     0.27} $  &     11& $^{+      0.00}_{-      8.56} $  &     13.08& $^{+      4.56}_{-      0.44} $&0.81  \\                     NFW&    200.00& $- $  &      0.62& $- $ &     63.70&    2.17& $- $  &      5.66& $- $  &     12.63& $-  $&0.40  \\
                   piso&     38.1& $^{+     21.96}_{-     28.05} $  &      3.81& $^{+     30.39}_{-      0.97} $  &     63.70&    1.69& $^{+     0.68}_{-     0.30} $  &     10.95& $^{+      0.05}_{-      8.51} $  &     13.32& $^{+      3.99}_{-      0.69} $&0.81  \\\hline
\multicolumn{12}{c}{Maximum-disc model }\\
\hline
           Burkert&     57.84& $^{+     33.22}_{-     16.38} $  &      4.41& $^{+      1.79}_{-      1.12} $  &     63.70&    1.64& $^{+     0.38}_{-     0.27} $  &     11& $-$  &     13.08& $^{+      0.96}_{-      0.44} $&0.81  \\                NFW&    161.82& $- $  &      0.50& $- $  &     63.70&    1.32& $- $  &     11& $- $  &     12.63& $-  $ &0.81 \\                                piso&     38.10& $^{+     16.38}_{-     13.13} $  &      3.81& $^{+      1.82}_{-      0.82} $  &     63.70&    1.69& $^{+     0.30}_{-     0.28} $  &     11& $-$  &     13.32& $^{+      0.71}_{-      0.69}$& 0.81  \\
\hline
\multicolumn{12}{c}{Photometrical model }\\
\hline
           Burkert&     24.79& $^{+     2.36}_{-     5.33} $  &      21.00& $^{+      10.02}_{-      2.86} $  &     63.70&    2.18& $^{+     0.04}_{-     0.16} $  &    2.44& $- $  &     15.79& $^{+      1.71}_{-      3.16} $&0.18  \\                NFW&    112.55& $^{+     64.65}_{-     36.60} $  &      1.58& $^{+     1.14}_{-     0.69} $  &     63.70&    2.46& $^{+     0.23}_{-     0.20} $  &     2.44& $- $  &     12.63& $-  $ &0.18 \\                                piso&     11.07& $^{+     5.38}_{-     1.35} $  &      29.54& $^{+      6.36}_{-      13.09} $  &     63.70&    2.19& $^{+     0.26}_{-     0.07} $  &     2.44& $- $  &     12.63& $^{+      0.00}_{-      4.99}$& 0.18  \\
\hline
\end{tabular}
\end{center}
\end{table*}

\section{Discussion}\label{Discussion}

\subsection{On the formation scenario of UGC~1922}
The main goal of this study is to understand how such an unusual system as UGC~1922 has been formed. We propose and discuss several possible scenarios of the formation of a gLSB galaxy with counter-rotation.
\begin{enumerate}
\item The giant disc could have been formed because of the large radial scale and low central density of the dark halo of UGC~1922, as it was proposed for Malin~2 by \citet{Kasparova2014}. Our data do not contradict this statement as long as we obtained extremely large radial scale of the dark halo (see Table \ref{par}) comparable to that of Malin~2 \citep{Kasparova2014}.  According to \citet{Saburova2018}, dark halo radial scales exceeding 10~kpc occur rarely in disc galaxies. This might have happened because of peculiar environment at the formation stage of these galaxies.

Following the formation of the gLSB disc, a  galaxy  with the mass significantly lower than that of the gLSB disc, but  still quite massive -- with the mass of order of $10^{10}$\Msun containing both stars and gas counter rotating with respect to the main disc was accreted by the bulge.  The stars of the bulge mixed with the rotating stellar population of the intruder flattened the stellar velocity gradient in comparison to that of the ionized gas. The observed SED and spectrum in the centre of UGC~1922 can be fitted by two stellar components, one of which has younger age and can be a fossil of the accreted galaxy with ongoing star formation at the time of the merger. The fact that UGC~1922 is visible in FUV in the centre can also speak in favour of the presence of young stellar population there unless it is the manifestation of the UV-upturn in old metal-rich stellar populations firstly discovered by \citet{Code1979} in early-type galaxies, which can originate from hot horizontal branch stars  \citep[see e.g.]{Greggio1990}.

The derived gradient of the gas metallicity is in agreement with the ``normal'' evolution in the framework of the ``inside-out'' scenario which can provide an additional argument in favour of a non-catastrophic scenario. However, in such a case it is difficult to explain the presence of composite ionization mechanisms in the disc of the galaxy and the irregular structure on its NW side (see above). Probably the formation of a giant disc itself may not be the result of merging, but we cannot exclude the possibility of gravitational interaction of UGC~1922 with neighbour galaxies, which has not destroyed the giant disc but left the traces in its morphology.

\item A scenario proposed by \citet{Penarrubia2006} in which elliptical galaxy accretes a large number of dwarf gas-rich galaxies, which form a giant disc. Then, the irregular clumps on the NW-side of the disc can be associated with a remnant of one such intruder. However, we do not see any signs of the decline of the rotation curve at the outermost radius predicted by \citet{Penarrubia2006} in the available \HI data. The flat gas metallicity radial profile also speaks against the recent accretion of dwarf metal-poor galaxies. To assemble the disc mass of at least $2\times 10^{11}$\Msun one needs a large number of such encounters, which seems to be not very realistic if to take into account that the satellites should have almost the same angular momentum to form a disc instead of a spherical system. Thus, this scenario looks less probable for UGC~1922.

\item The slow accretion of gas from a filament on an elliptical galaxy. The high effective radius of the bulge and absence of stellar rotation in the centre of UGC~1922 makes it a slow rotator in the classification by \citet{Emsellem2011}, which can speak in favour of the scenario. The Sersic index of the bulge is about 2 in contrast to 4 expected for ellipticals, and the accretion should have taken place shortly after the formation of an elliptical galaxy (within 5 Gyr) because the age of the stellar disc is not significantly younger than that of the bulge.\footnote{We do not cover the pure disc region by our spectroscopic observations, however red colour of the diffuse emission  at the outskirts of the disc of UGC~1922 $g-r\sim0.6-0.8$ favours the presence of old stellar population.} The metallicity of the ionized gas and stars in the disc is not dramatically lower than that in the centre and could be explained by a typical metallicity gradient observed in disc galaxies of this surface brightness. An alternative is that the accreted gas should be enriched by heavy elements, which questions its origin from a cosmic filament.

\item The result of a merger of a giant Sa galaxy and a gas rich giant Sd companion on a prograde co-planar orbit (inclination of 0\degr) with the initial and pericentral distances of 100~kpc and 24~kpc respectively. According to tree-SPH simulations extracted from the {\sc galmer} database \citep{Chilingarianetal2010}, such a merger scenario within 3~Gyr will form a system with a bulge surrounded by a giant disc with spiral arms with the radius exceeding 100~kpc. The gas in the resulting galaxy will be concentrated in spiral arms (see the {\sc galmer} database, model 559 -- the result of a merger between gSa and gSd using the orbit type 9). The presence of irregular clump on the NW-side of the disc of UGC~1922 could be a trace of a major merger in this case.

However, the central counter-rotation is not reproduced in this model. Thus, we still need to incorporate a minor merger that formed a kinematically decoupled inner disc.  According to {\sc galmer} database the mass of the in-falling galaxy could be as high as $10^{10}$\Msun, which makes our formation scenario resembling that proposed by \cite{Zhu2018}, since it also incorporates the merging of three quite massive systems.  If one considers  the metallicity of stars of the young central component to be solar (see Sect. \ref{2c}) and that the gas metallicity radial gradient unfavours a recent accretion of metal-poor gas in the centre, a minor merger that could have led to the decoupled central kinematics must have taken place a long time ago. It could have happened even before the major merger. However, it is important to know whether the counter-rotation can be preserved after the catastrophic interaction.
\end{enumerate}

\subsection{High resolution simulations of a Sa+Sd merger}

In order to find out whether the counter-rotation can be preserved after a major merger of two massive galaxies with a pre-existing counter-rotating component, we performed dedicated $N$-body magnetohydrodynamics~(MHD) simulations of a merger of two disc galaxies. The simulation was performed with our fully three-dimensional code described in ~\citet{2014JPhCS.510a2011K} which has been used in a number studies~\citep{2015MNRAS.451.2889K,2017MNRAS.468..920K,2018A&A...609A..60K}. The MHD part is treated by using TVD MUSCL scheme with HLLD solver and constrained transport for the magnetic field divergence cleaning. Gravity is solved using a binary tree, and N-body system integrated using a kick-drift-kick leapfrog. Self-gravity is active for all components, using a fixed gravitational softening of $50$~pc. Except for the DM, stellar and gas-dynamics, we also introduce star formation, stellar chemical evolution, and ISM magnetic field treatment. We also take into account an approximation for cooling/heating processes for ISM of solar metallicity~\citep{2013MNRAS.428.2311K,2016MNRAS.455.1782K}.

We model the host disc galaxies as a self-consistent three-component system, containing a dark matter halo, stellar and gaseous discs. The halo is used in a form of an isothermal sphere, while stellar and gaseous discs are taken in Myamoto--Nagai approximation. Equilibrium of the galaxies is constructed following the procedure outlined by~\citet{2009MNRAS.392..904R}. The chosen values for the parameters of both disc galaxies and orbital parameters were taken from the model 559 presented in {\sc galmer} database~\citet{Chilingarianetal2010},  with the exception of the gaseous components, see below (for more details on the {\sc galmer} model see above). We use a total number of particles of 12 millions, corresponding to 6 million particles per galaxy (stars and dark matter). Spatial resolution for the gaseous component is constant and equals to $\approx 90$~pc. The initial setup was chosen to be representative of a wet merger with a gas fraction of $10$ and $1.5$ for the host galaxies.

The key difference between our model and {\sc galmer} model is the following. We change the direction of spin of gaseous and stellar components in Sa galaxy within $5$~kpc by making its rotation in the same plane but in the opposite direction without spatial overlap between the components. Such initial conditions allow us to study the survival of the counter-rotating component via major merger.

The initial disc galaxies are on a parabolic orbit and lose angular momentum each other via resonant gravitational torques and sink to the mass center. Globally time-depended evolution is very much similar to the {\sc galmer} simulation. During the first passage torques exerted by the systems drive some gas to the galaxies centers and introduce strong shocks in tidal and `bridge' regions between the galaxies, which drives a burst of star formation~\citep[see e.g.,][]{2007A&A...468...61D,2010ApJ...720L.149T,2015MNRAS.446.2038R}.

\begin{figure*}
\includegraphics[width=0.25\linewidth]{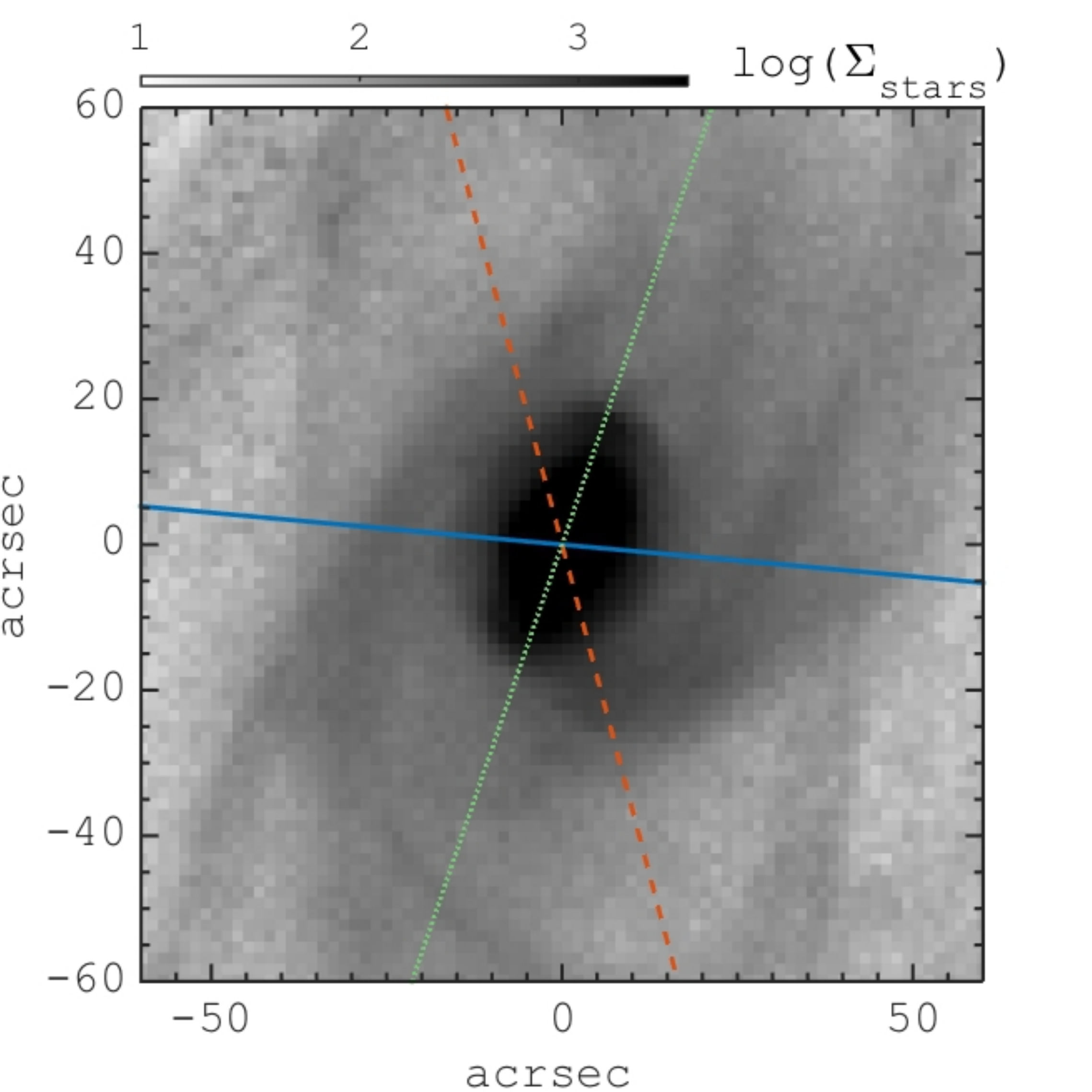}\includegraphics[width=0.25\linewidth]{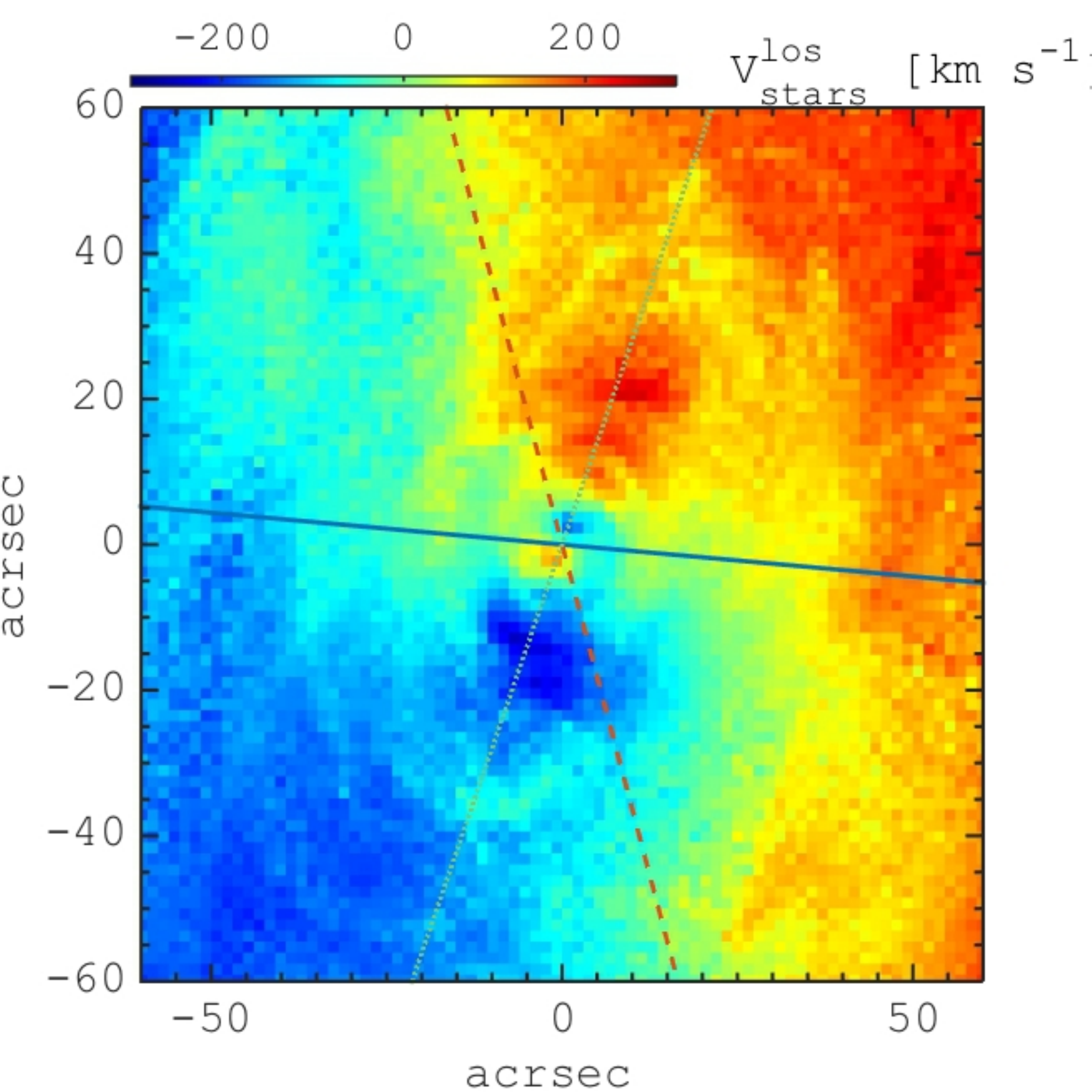}\includegraphics[width=0.25\linewidth]{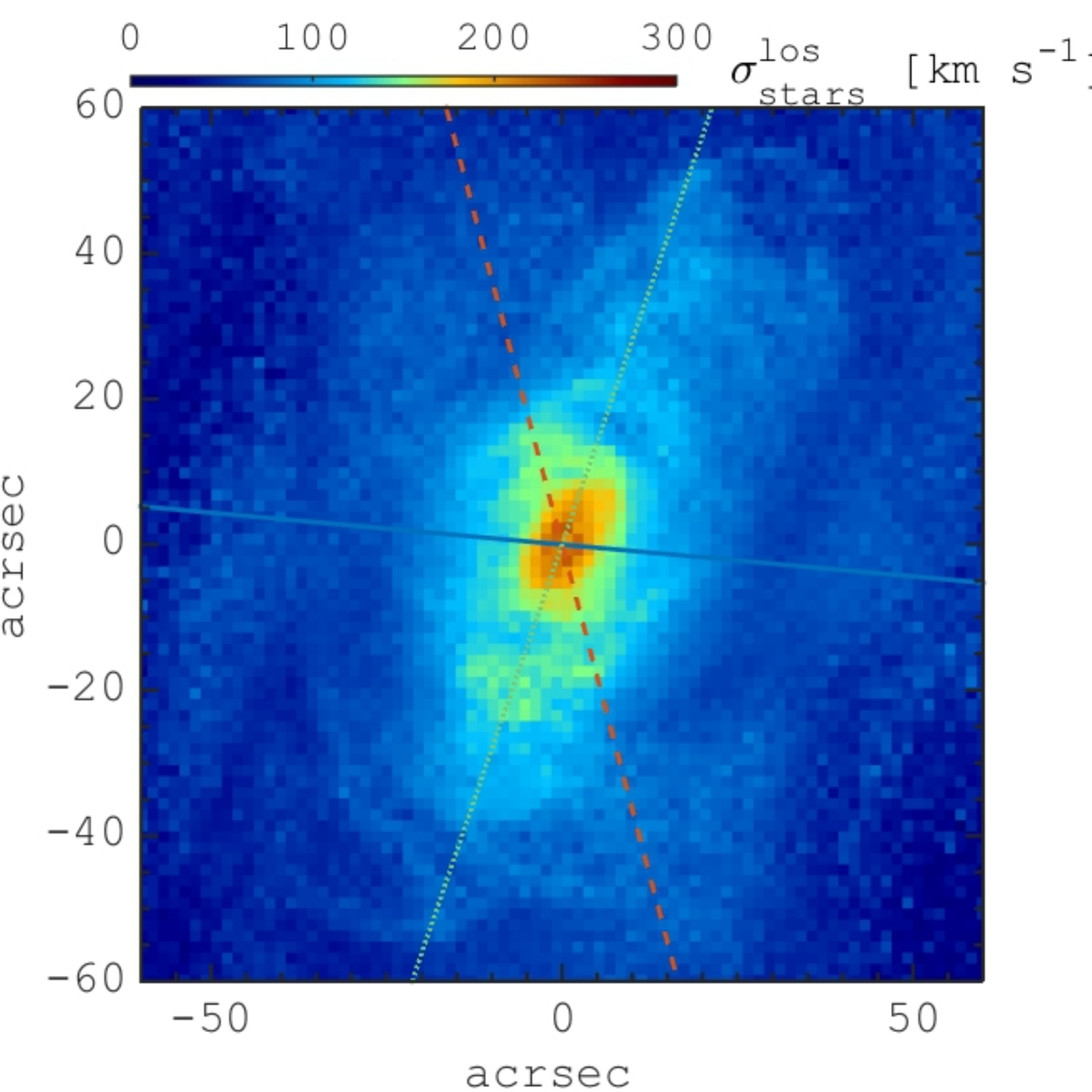}\includegraphics[width=0.25\linewidth]{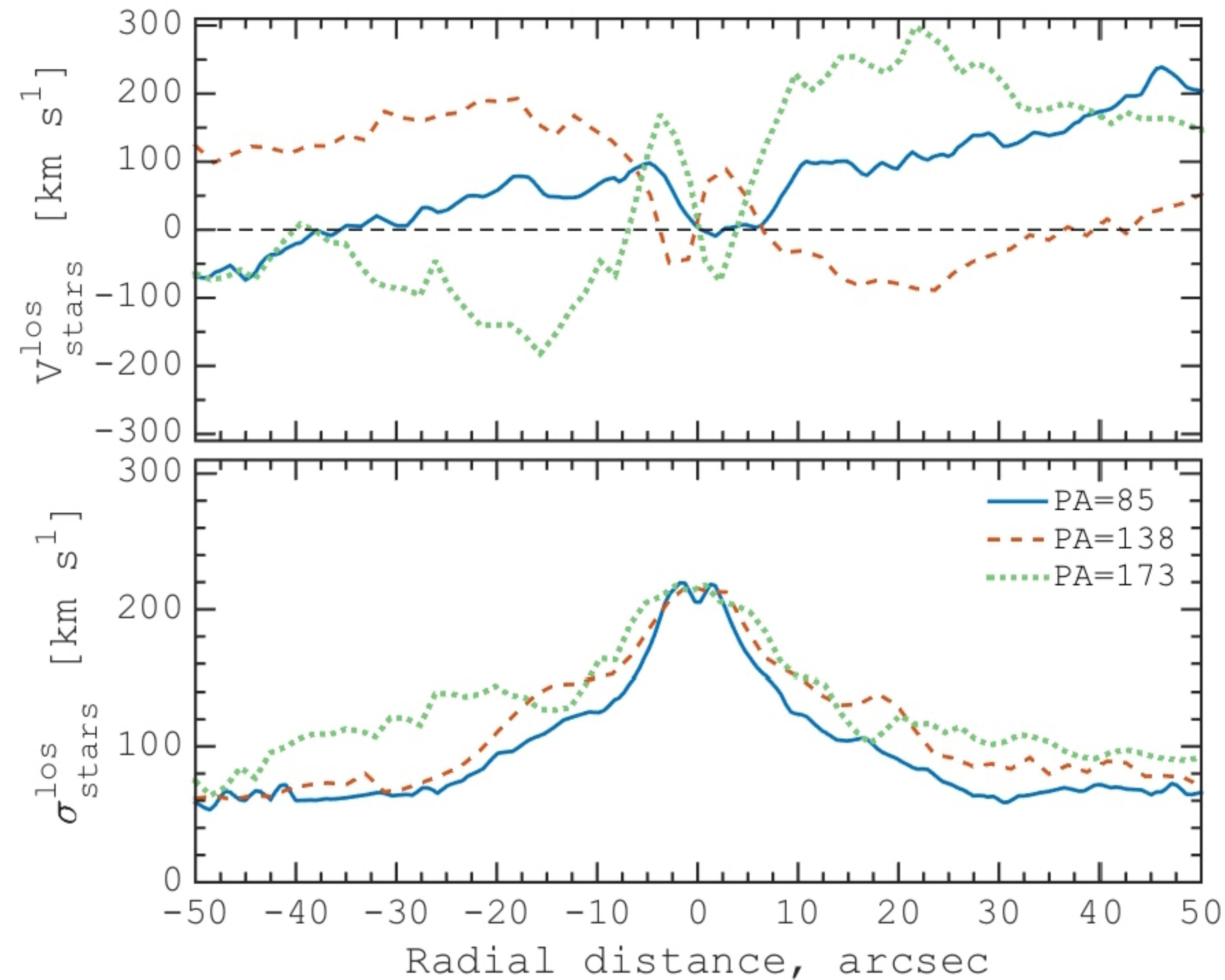}\\\includegraphics[width=0.25\linewidth]{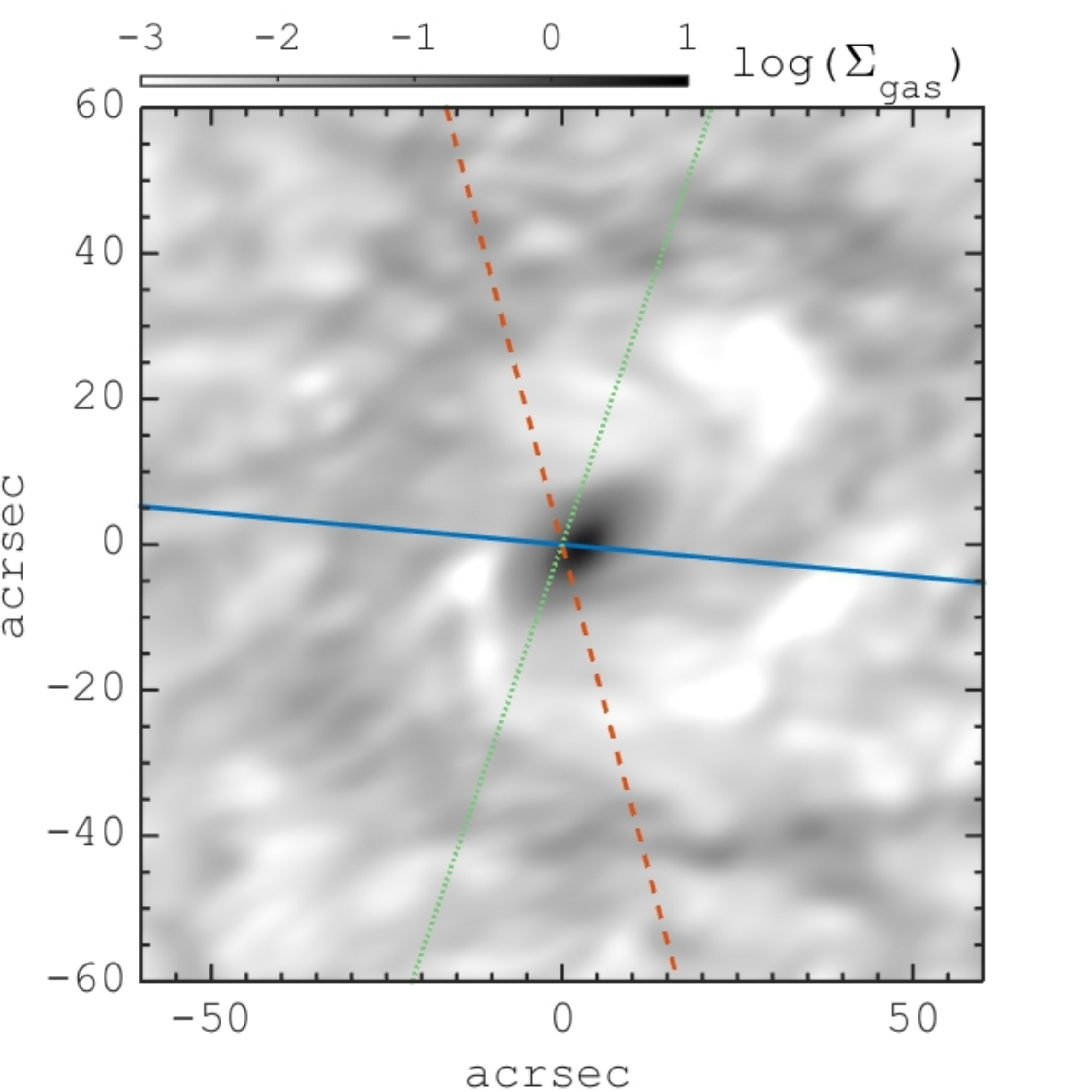}\includegraphics[width=0.25\linewidth]{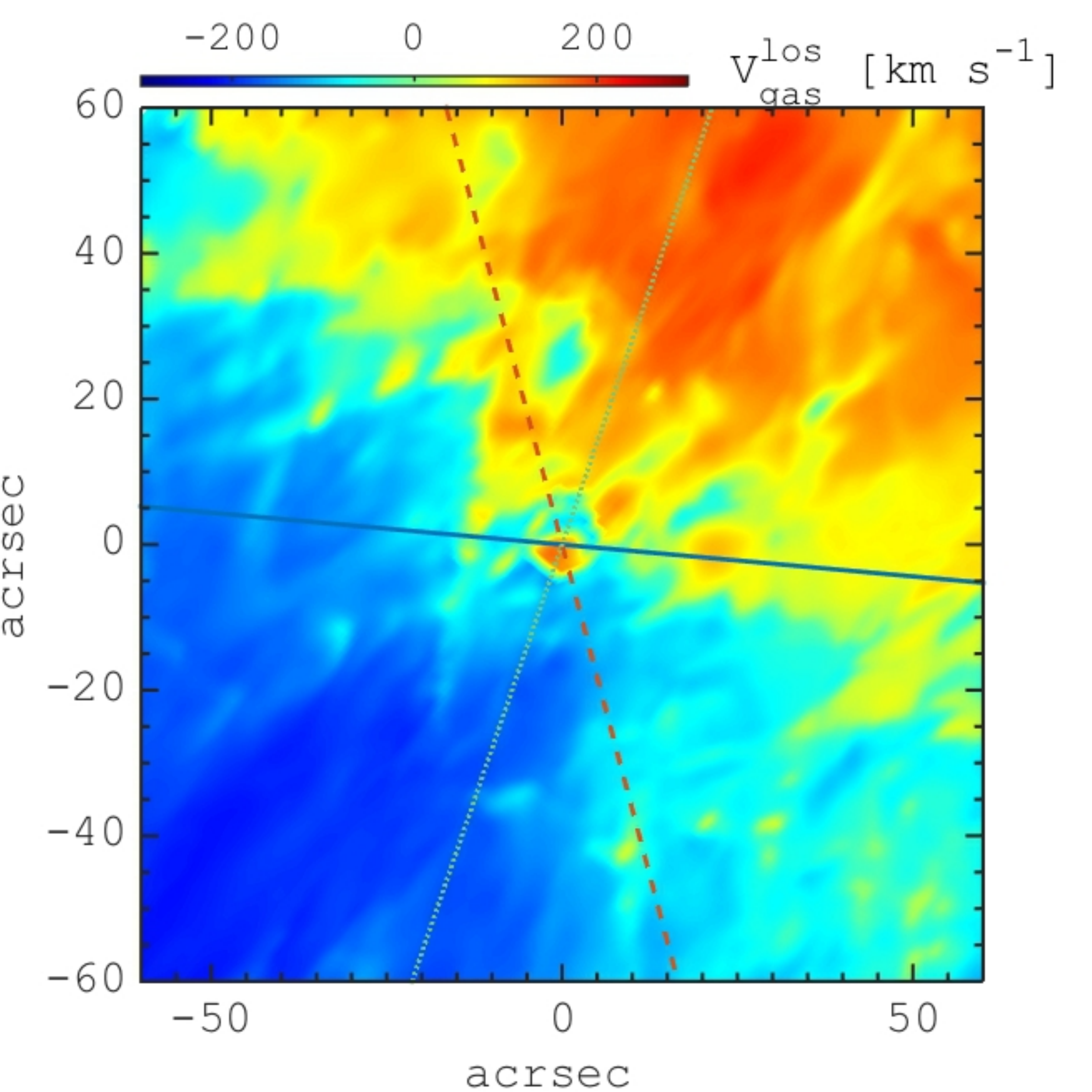}\includegraphics[width=0.25\linewidth]{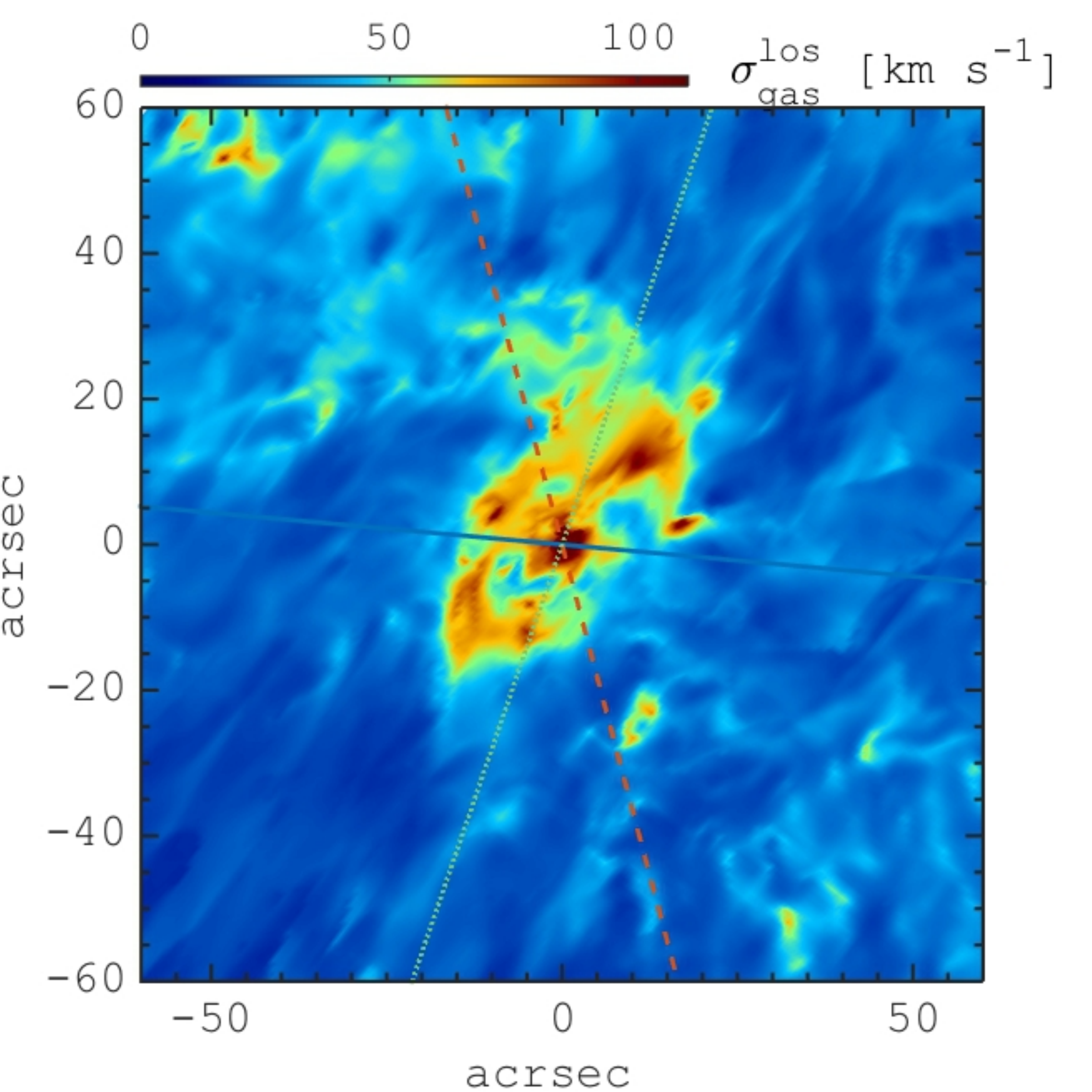}\includegraphics[width=0.25\linewidth]{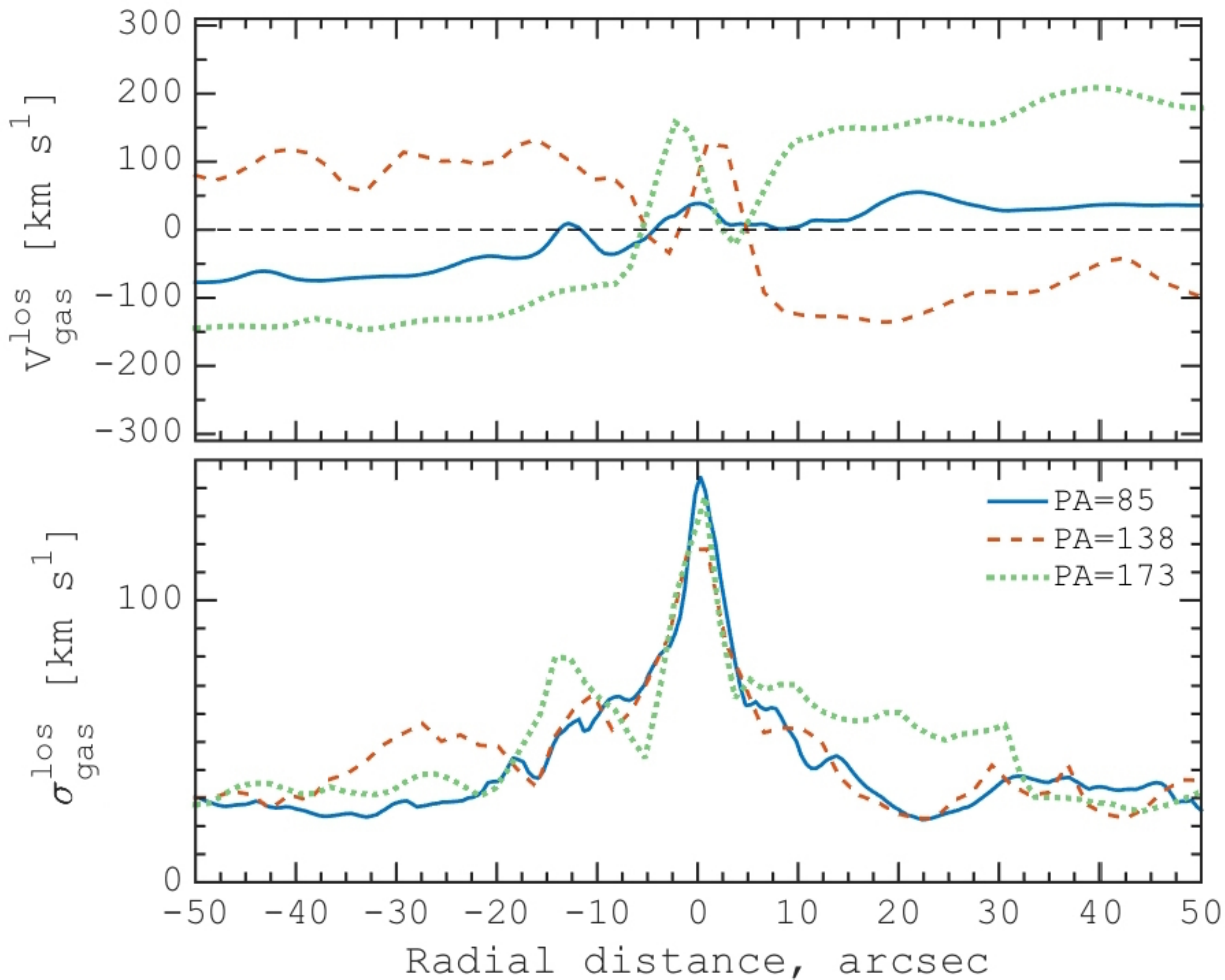}\caption{Result of numerical simulation of two disc galaxy merger after $2.5$~Gyr of evolution. Resulting galaxy is orientated to mimic the observed spatial orientation of UGC~1922. {\it Top:} from left to right: stellar projected density, LOS velocity~(upper frame) and LOS velocity dispersion~(lower frame) as well as kinematics of stars along three position angles mentioned in the label. {\it Bottom:} the same parameters, but for gaseous component. Maps and one-dimensional data are plotted with uniform spatial resolution of $1.5$~arcsec.}
\label{fig::models}
\end{figure*}

After $2.5$ Gyrs of evolution the system is relaxed in the central part~($\leq 30-40$~kpc)  while tidal spiral arms are still visible around it. Since we simulate an in-plane merger, the remnant has the shape, profile and kinematics of largely extended disc galaxy~(see Fig.~\ref{fig::models}, left frames). At the end of the simulation, the morphology resembles that observed in UGC 1922~(see Fig.~\ref{map}). Note also that a few compact clumpy structures (mostly containing young stars) are visible across the entire disc galaxy as in UGC~1922. Star formation and chemical evolution model allows us to trace the distribution of young stars. We found that younger~(and more metal-rich) stellar populations are distributed along the spiral structures~(seen in~Fig.~\ref{fig::models}, top left) and in the galactic center~($r<2-3$~kpc) similar to UGC~1922.

Kinematics of the merger remnant is very interesting, in fact, its stellar component depicts counter-rotation within $r<5$  kpc which is similar to those we set up initially for one of our host galaxies. Despite the overall galaxy kinematics is quite complex, for the gaseous component~(see bottom line in Fig.~\ref{fig::models}) we also see a counter-rotating feature in the center. Global LOS velocity maps (and profiles) of gas and stars are very much similar across the galaxy. Velocity dispersion exhibits an increase in the central region which  partially can be enhanced by the action of counter-rotating component~(see details in~\citealt{2017A&A...597A.103K}).  

In the end, our model shows that the counter-rotation preserves in both stellar and gaseous components. At the same time, the morphology of the model galaxy, the distribution of young metal-rich stars, the high central velocity dispersion of gas and stars, the stellar kinematics are in good agreement with what we observe in UGC~1922. 

Since our model do reproduce the counter-rotation in the galactic center, we believe that the major merger of galaxies with different initial gas fraction, one of which has already undergone merger with a gas-rich galaxy with counter-rotation and with total mass of order of $10^{10}$ solar masses is a reasonable scenario for the formation of UGC~1922-type galaxies. This scenario resembles that proposed by \citet{Zhu2018} for the formation of Malin~1-type galaxy, which also involves the merging of three galaxies. However, a more precise understanding of the galaxy formation requires a detailed parameter study via high-resolution $N$-body/hydrodynamical simulations of galaxy mergers.

\subsection{Clues to the formation history of gLSBGs}
Despite the fact that we can not exclude the scenarios (i) and (iii) listed above for the formation of UGC~1922, we give more preferences to the variant (iv) since it does not contradict the observational data reproduced in our $N$-body simulations. 

The most important question is if the suggested formation scenarios can be applied to gLSBGs in general. The scenario (i) has its advantage that it can give us clues on the formation of giant discs in general. At the same time the parameters of merger in scenario (iv) make it a very rare event, which possibly can not explain the formation of all gLSBGs if more representatives of this unusual class are found. 

We suspect that in the case of gLSB discs, the environment can play important role both at the stage of the formation and in the more recent epoch -- these systems should avoid dense environment in order to preserve their discs while one also needs special conditions to form both massive and rarified dark halos. In fact, despite UGC~1922 shares common morphological features with other gLSBGs such are spiral arms and prominent bulge,  the environment of UGC~1922 differs from that of most of other known galaxies of this class. As we describe above UGC~1922 belongs to a group that includes 7 spectroscopically confirmed members, one of which is a giant elliptical galaxy that dominates the group. At the same time, UGC~1382 is situated in low-density environment possibly belonging to a very poor group with only 2 members \citep{Hagen2016}. Malin~2 has four low-massive companions \citep{Kasparova2014}. Malin~1 is also located in a relatively low-density region, near the structure that could be attributed to the filament of the large scale structure. It possesses low-massive close companion and bright galaxy 350 kpc far from it with the velocity difference of roughly 100 \kms \citep{Reshetnikov2010}. UGC~6614 is one of three members of the group \citep{Crook2007}. NGC~7598 also belongs to poor group with only two members \citep{Saulder2016}. 

Therefore UGC~1922 is not typical in this respect. However, if we consider the deep images of other gLSBGs, we will see that besides spiral arms they  have  diffuse structures in the discs similar to the NW cloudy structure that we observe in UGC~1922  \citep[see, e.g.]{Boissier2016, Kasparova2014,Hagen2016}. It can indicate the presence of interaction in the history of the systems. In particular, the Sa+Sd in-plane merger that we consider here can potentially create such structures. Thus, we can not exclude that the scenarios proposed for UGC~1922 are also valid for other gLSBGs despite the difference in the environment. However, more data, especially on inner kinematics and the radial variation of metallicity in gLSBGs are needed to test this hypothesis. 

\section{Conclusions}\label{conclusion}

We present the results of long-slit spectroscopic and photometric observations of a Malin~1 ``cousin'', the giant LSB galaxy UGC~1922 that has a {\it g}-band disc radius of 84~kpc. The data reveal that:
\begin{itemize}
\item Gas and stars in central part of UGC~1922 show opposite gradient of the velocity with respect to that of the main disc of the galaxy.
\item The age and metallicity of stars in the LSB disc of UGC~1922 is lower than in central part, so is the metallicity of the gas. 
\item The gas metallicity radial gradient is in agreement with that found for moderate-size LSB-galaxies \citep{Bresolin2015}. 
\item The difference of metallicity of gas and stellar age in the centre and in the disc is not dramatic, so we can rule out the formation of the giant LSB disc from the metal poor gas accreted from the filament.
\item We propose that the following formation scenarios could explain the observed properties of the galaxy. The first scenario is catastrophic and more preferable. According to it the giant LSB disc was formed by the merger of massive Sa and Sd galaxies with zero inclination with respect to Sa disc and prograde orbit.   In the second non-catastrophic scenario UGC~1922 could be formed due to the high radial scale and low central density of the dark halo, which is not in conflict with our observational data.  In both scenarios the counter-rotation in the centre could origin from the accretion of  gas-rich galaxy. According to the results of our $N$-body simulations the stellar  and gaseous counter-rotation will preserve in the catastrophic scenario even if the gas-rich galaxy was accreted by one of the galaxies before the merging.
\item The upper limit of the  mass of stellar disc obtained from the velocity dispersion of stars in the assumption of the disc marginal gravitational stability appears to be five times higher than follows from the observed colour index and model relations of \citet{Roediger2015}. The high disc mass-to-light ratio is in the better agreement with baryonic Tully-Fisher relation and is not in conflict with observed rotation curve. It  indicates that UGC~1922 could contain significant amount of dark matter in baryonic or non-baryonic form in the disc.
\end{itemize}

\section*{Acknowledgements}
The authors thank the anonymous referee for valuable comments. 
AS is grateful to Anatoly Zasov for fruitful discussion. We thank Alka Mishra for providing \HI data.
The observations at the 6-meter BTA telescope were carried
out with the financial support of the Ministry of Education
and Science of the Russian Federation (agreement No. 14.619.21.0004, project ID RFMEFI61914X0004).
We used the equipment purchased
through the funds of the Development Program of
the Moscow State University.
The spectral data reduction and interpretation of the results were supported by The Russian Science Foundation (RSCF) grant No. 17-72-20119. The $N$-body/hydrodynamical simulations were carried out using the equipment of the shared research facilities of HPC computing resources at Lomonosov Moscow State University supported by the project RFMEFI62117X0011 and RFBR project 16-02-00649. SAK has been supported by RFBR grant 16-32-60043. IC's research is supported by the Telescope Data Center of the Smithsonian Astrophysical Observatory. We acknowledge the usage of the HyperLeda database (http://leda.univ-lyon1.fr). Based on observations obtained with MegaPrime/MegaCam, a joint project of CFHT and CEA/DAPNIA, at the Canada-France-Hawaii Telescope (CFHT) which is operated by the National Research Council (NRC) of Canada, the Institut National des Science de l'Univers of the Centre National de la Recherche Scientifique (CNRS) of France, and the University of Hawaii.
The Pan-STARRS1 Surveys (PS1) and the PS1 public science archive have been made possible through contributions by the Institute for Astronomy, the University of Hawaii, the Pan-STARRS Project Office, the Max-Planck Society and its participating institutes, the Max Planck Institute for Astronomy, Heidelberg and the Max Planck Institute for Extraterrestrial Physics, Garching, The Johns Hopkins University, Durham University, the University of Edinburgh, the Queen's University Belfast, the Harvard-Smithsonian Center for Astrophysics, the Las Cumbres Observatory Global Telescope Network Incorporated, the National Central University of Taiwan, the Space Telescope Science Institute, the National Aeronautics and Space Administration under Grant No. NNX08AR22G issued through the Planetary Science Division of the NASA Science Mission Directorate, the National Science Foundation Grant No. AST-1238877, the University of Maryland, Eotvos Lorand University (ELTE), the Los Alamos National Laboratory, and the Gordon and Betty Moore Foundation.

\bibliographystyle{mnras}
\bibliography{saburova}

\label{lastpage}

\end{document}